\documentclass[prb,preprintnumbers,showpacs,twocolumn,amsmath,amssymb,superscriptaddress]{revtex4-1}
\usepackage{color}
\usepackage{times}
\usepackage{epsf}
\usepackage{float}
\usepackage{graphicx}% Include figure files
\usepackage{epsfig}
\usepackage{epstopdf}
\usepackage{dcolumn}% Align table columns on decimal point
\usepackage{bm}% bold math

\def\<{\langle}
\def\>{\rangle}
\def\({\left(}
\def\){\right)}
\def\[{\left[}
\def\]{\right]}

\def\cos{\mathop{\mathrm{cos}}\nolimits}

\def\aa{\textbf{\textit{a}}}

\def\kk{\textbf{\textit{k}}}

\def\ww{\textbf{\textit{w}}}

\def\rr{\textbf{\textit{r}}}
\def\uu{\textbf{\textit{u}}}

\def\ee{\textbf{\textit{e}}}

\def\UU{\textbf{\textit{U}}}

\def\RR{\textbf{\textit{R}}}

\def\XX{\textbf{\textit{X}}}

\begin{document}
\draft

\title{Hopping Processes Explain {\it T}-linear Rise of Thermal Conductivity in Thermoelectric Clathrates above the Plateau}
\author{Qing Xi}
\affiliation{Center for Phononics and Thermal Energy Science,
School of Physics Science and Engineering, Tongji University, 200092
Shanghai, P. R. China}
\affiliation{China-EU Joint Center for Nanophononics, School of Physics Science and Engineering, Tongji University, 200092 Shanghai, P. R. China}
\affiliation{Shanghai Key Laboratory of Special Artificial Microstructure Materials and Technology, School of Physics Science and Engineering, Tongji University, 200092 Shanghai, P. R. China}

\author{Zhongwei Zhang}
\affiliation{Center for Phononics and Thermal Energy Science,
School of Physics Science and Engineering, Tongji University, 200092
Shanghai, P. R. China}
\affiliation{China-EU Joint Center for Nanophononics, School of Physics Science and Engineering, Tongji University, 200092 Shanghai, P. R. China}
\affiliation{Shanghai Key Laboratory of Special Artificial Microstructure Materials and Technology, School of Physics Science and Engineering, Tongji University, 200092 Shanghai, P. R. China}

\author{Jie Chen}
\affiliation{Center for Phononics and Thermal Energy Science,
School of Physics Science and Engineering, Tongji University, 200092
Shanghai, P. R. China}
\affiliation{China-EU Joint Center for Nanophononics, School of Physics Science and Engineering, Tongji University, 200092 Shanghai, P. R. China}
\affiliation{Shanghai Key Laboratory of Special Artificial Microstructure Materials and Technology, School of Physics Science and Engineering, Tongji University, 200092 Shanghai, P. R. China}

\author{Jun Zhou}
\email{zhoujunzhou@tongji.edu.cn}
\affiliation{Center for Phononics and Thermal Energy Science,
School of Physics Science and Engineering, Tongji University, 200092
Shanghai, P. R. China}
\affiliation{China-EU Joint Center for Nanophononics, School of Physics Science and Engineering, Tongji University, 200092 Shanghai, P. R. China}
\affiliation{Shanghai Key Laboratory of Special Artificial Microstructure Materials and Technology, School of Physics Science and Engineering, Tongji University, 200092 Shanghai, P. R. China}

\author{Tsuneyoshi Nakayama}
\email{tnaka@eng.hokudai.ac.jp}
\affiliation{Center for Phononics and Thermal Energy Science,
School of Physics Science and Engineering, Tongji University, 200092
Shanghai, P. R. China}
\affiliation{China-EU Joint Center for Nanophononics, School of Physics Science and Engineering, Tongji University, 200092 Shanghai, P. R. China}
\affiliation{Shanghai Key Laboratory of Special Artificial Microstructure Materials and Technology, School of Physics Science and Engineering, Tongji University, 200092 Shanghai, P. R. China}
\affiliation{Hokkaido University, 060-0826
Sapporo, Japan}

\author{Baowen Li}
\email{Baowen.Li@colorado.edu}
\affiliation{Department of Mechanical Engineering, University of Colorado, Boulder, Colorado 80309, USA}

\date{\today}

\begin{abstract}
Type-I clathrate compounds with off-center guest ions realize the phonon-glass electron-crystal concept by exhibiting almost identical lattice thermal conductivities $\kappa_{\rm L}$ to those observed in network-forming glasses.
This is in contrast with
type-I clathrates with on-center guest ions showing  $\kappa_{\rm L}$ of conventional crystallines.
Glasslike $\kappa_{\rm L}$ stems from the peculiar THz frequency dynamics in off-center type-I clathrates where there exist three kinds of modes
classified into extended(EX), weakly(WL) and strongly localized(SL) modes as demonstrated by Liu $et.\,al.$, Phys. Rev. B {\bf 93}, 214305(2016).
Our calculated results based on the hopping mechanism of SL modes via anharmonic interactions show fairly good agreement with observed {\it T}-linear rise of $\kappa_{\rm L}$ above the plateau.
We  emphasize that both the magnitude and the temperature dependence
are in accord with the experimental data of off-center type-I clathrates.
\end{abstract}

\maketitle
%\newpage
\section{INTRODUCTION}
Lattice thermal conductivity constitutes a key element to improve the efficiency of the thermal-to-electrical conversion in thermoelectric (TE) devices as understood from the material's figure of merit describing the efficiency $Z=S^2\sigma/\kappa_{\rm tot}$\,[K$^{-1}$].
The numerator contains the Seebeck coefficient $S(T)\,$[V/K] and the electrical conductivity $\sigma(T)\,$[1/($\Omega$m)], while
the denominator $\kappa_{\rm tot}(T)$\,[W/(mK)] consists of the sum of electrical $\kappa_{\rm el}$ and lattice   $\kappa_{\rm L}$ thermal conductivity.
Hence, the high performance of thermoelectricity can be achieved for materials with the lowest possible thermal conductivity $\kappa_{\rm tot}$, the highest possible electrical conductivity $\sigma$ and the highest possible Seebeck coefficient $S$.
Provided that the Wiedemann-Franz law $\kappa_{\rm el}(T)\propto\sigma(T)$ holds for, $\kappa_{\rm L}$ becomes a crucial parameter to improve the performance of TE conversion.
In this framework, Slack~\cite{Slack1995} has proposed the concept of ``phonon-glass electron-crystal".
This has been one of guiding principles for exploring high-performance TE materials~\cite{Takabatake2014, Beekman2015}.

Type-I clathrates with ``off-center" guest ions, such as R$_8$Ga$_{16}$Ge$_{30}$(R=Ba, Sr, Eu)\,\cite{Nolas:1998aa, Cohn:1999a, Christensen:2016a,Paschen:2001a,Sales:2001a, Avila:2006a}, Ba$_8$Ga$_{16}$Sn$_{30}$\,\cite{Avila:2008aa, Suekuni:2008a}, Sr$_8$Ga$_{16}$Si$_{30-\mathrm{x}}$Ge$_{\mathrm x}$\,\cite{Suekuni:2007a}, are particularly interesting in this respect since these systems exhibit almost identical lattice thermal conductivities to those of structural glasses, which consist of four specific regions characterized by: (i)\,{\it T}$^{\sim 2}$-dependence below a few Kelvin, (ii)\,the plateau region between a few K and a few 10K, and (iii)\,the subsequent rise proportional to {\it T}, and (iv)\,its saturation above {\it T}$\sim$100K.
These characteristics of $\kappa_{\rm L}$ exhibit a remarkable uniformity which appears to be insensitive to chemical compositions, suggesting the existence of a unified mechanism~\cite{Nakayama2002}.
However, this issue remains as an open and challenging problem of long-standing due to the difficulty to identify relevant entities or elements at atomistic level caused by their complex microscopic structures.
%Sr$_8$Ga$_{16}$Ge$_{30}$\,\cite{Nolas:1998aa, Cohn:1999a, Christensen:2016a}, Eu$_8$Ga$_{16}$Ge$_{30}$\,\cite{Paschen:2001a},
Surprisingly enough, though ``off-center" clathrates are crystalline with regularly network structure, the temperature dependence as well as the magnitudes of their thermal conductivities are almost identical to those of structural glasses over the full temperature range.
In contrast, type-I clathrates with ``on-center" guest ions show conventional crystalline $\kappa_{\rm L}$~\cite{Takabatake2014}.

This paper is organized as follows.
Section II surveys the characteristics of vibrational modes according to the results of the spectral density of states, eigenvalues and their eigenvectors~\cite{Liu2016}.
We claim in this Section that the onset of the plateau is due to the delocalization-localization (weak localization) transition of acoustic modes.
In addition, we point out that the temperature region showing the {\it T}-linear rise subsequent to the plateau is associated with the energy range where SL modes are fully excited.
Section III
describes the construction of anharmonic interaction
Hamiltonian between SL and EX modes.
The second quantized form of anharmonic Hamiltonian is given in Section IV.
Section V develops a theory on the mechanism governing the {\it T}-linear rise of $\kappa_{\rm L}(T)$ above a few 10\,K.
Excited modes in this temperature region are mostly strongly-localized (SL) modes satisfying the Ioffe-Regel condition as evident from the mode pattern obtained by large-scale numerical simulations~\cite{Liu2016}.
% characterized by that those wavelengths are of the order of the localization length for strongly localized moes.
These are hybridized modes between acoustic phonons associated with network cages and local vibrations of guest ions in cages.
%The modes possess the nature of ``optic" modes vibrating in out-of-phase between cages and guest ions.
Based on these numerical evidences, we explain in  quantitative manner $\kappa_{\rm L}(T)$ proportional to {\it T} observed above the plateau, by introducing the quantum mechanical process of hopping of SL modes due to anharmonic interactions, first proposed for fracton excitations\,\cite{Alexander1987}.
%Hereafter we call this process as ''corn-on-hotpan process".
%The mechanism is the SL mode version of the Mott's phonon-assisted hopping of localized electrons\,\cite{Mott1969}, which has been employed for heat transport in amorphous materials\,, amorphous materials and silica aerogels\,\cite{Jagannathan1989}.
%These works has claimed that the onset of the plateau in $\kappa_{\rm L}$ is concerned with the direct crossover from EX acoustic phonons to SL modes, not from EX modes to WL modes as demonstrated in Ref.\,\cite{Liu2016}.
%, and for structural glasses% by Nakayama and Orbach\,\cite{Nakayama1999gg}.
%In addition, the difficulty to identify the microscopic structure in these materials has limited theories within phenomenological ones, while the situation for off-center type-I clathrates is more clear-cut than the case of amorphous materials categorized into complex disordered systems\,\cite{Liu2016}.
%It will be shown that our formula  satisfactorily explains the magnitude and the temperature dependence of $\kappa_{\rm L}(T)$  observed  above the plateau.
%At first, we introduce the anharmonic interaction between extended acoustic modes and strongly localized modes, which are hybridized-modes between acoustic modes involved in network cages and local vibrations of guest ions.
%The second quantized Hamiltonian enables us to treat the hopping of strongly localized modes assisted by extended acoustic modes.
Summary and conclusions are given in Sec.\,VI.
\begin{figure}[t]
%\epsfysize=1.0in
%\epsfbox{anharmonic_potential}
\includegraphics[width=1.0\linewidth]{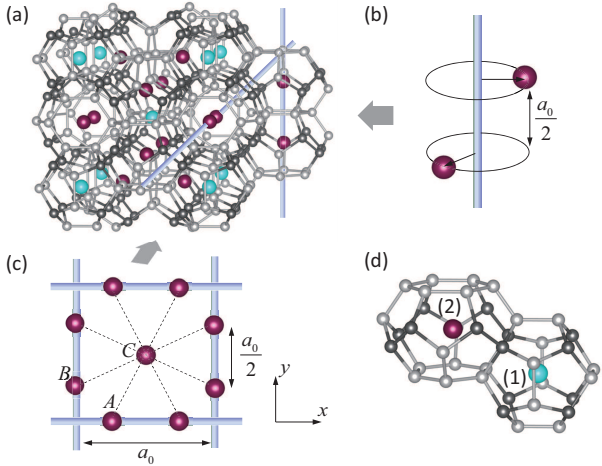}
\caption{(Color online)
\label{fig1}
(a)\,Illustration of type-I clathrate.
The fourfold inversion axes are directed along the axises $x, y, z$.
Red and blue balls represent off-center guest ions in tetrakaidecahedron cages and centered guest ions in dodecahedron cages, respectively.
(b)\,Two off-center guest ions along $y$-axis  are depicted.
(c)\, The configuration
of eight nearest neighbor guest ions connected by equilateral triangle. The sites A, B, and C in (c)
are seated on the chains parallel to $x, y,$ and $z$, respectively:
A=(a/4, 0, a/4), B = (0, a/4, 3a/4), C =(a/2, a/2, a/2).
%The distance between the next-nearest neighbors (dashed lines) is 3/8 a.
(d)\,The molecular unit composed of tetrakaidecahedron cage with off-center guest ion(2) at 24k site and smaller dodecahedral cage with guest ion(1) at 2a site.
%(f) The 3D configuration of chains.
}
\end{figure}
%%%%%%%%%%%%%%%%%%%%%%%%%%%

%%%%%%%%%%%%%%%%%%%%%%%%%
\section{CHARACTERISTICS OF EXCITED PHONONS AT THz FREQUENCY REGION}
%%%%%%%%%%%%%%%%%%%%%%%%%%%%%%%5
Type-I clathrates form a primitive cubic structure ($Pm\bar{3}n$) consisting of  6 tetrakaidecahedron (14-hedrons) and 2 dodecahedron (12-hedrons) per unit cell, in which the group-I or -II elements in the periodic table are encaged in the polyhedrons as guest ions. See Fig.\,\ref{fig1}.
The THz frequency phonon dynamics of off-center type-I clathrates has been investigated in terms of large-scale numerical simulations.
% for the system sizes of 20$\times$20$\times$20$\sim$100$\times$100$\times$100  by Liu $et.\,al.$~\cite{Liu2016}.
They have
%treated 10$^7\times 10^7$ sizes of matrices by
illustrated type-I Ba$_8$Ga$_{16}$Sn$_{30}$ \,(BGS) exhibiting glasslike $\kappa_{\rm L}(T)$ as a prototype material with off-center guest ions, in which the guest ion Ba(2) in tetrakaidecahedron cage has the  mass $m$ and the molecular unit
composed of one tetrakaidecahedron and 1/3 dodecahedron does the total mass $M$ excluding the off-center guest ion.
The coarse-grained picture, an operation
of reducing the degrees of freedom of the original system, is valid for our purpose from the following reasons.
First of all, EX acoustic modes at THz frequencies play a dominant role in heat transport since optical modes concerning to the vibrations of cages themselves do not contribute to thermal conductivity.
Second, the wave-length $\lambda$ of phonons in the frequency regime $\nu$\,$\leq$\,2.5\,THz ($E$\,$\leq$\,10\,meV) becomes $\lambda$\,$\geq$\,1.6 nm, which is larger than the size of a unit cell of $a_0$\,$\simeq$1\,nm in type-I clathrates, as estimated from the relation $\lambda$=$v/\nu$ using the sound velocity $v\approx 4\times10^3$[m/sec].
These validate the coarse-grained Hamiltonian for describing THz frequency dynamics rather than treating all microscopic constituents as equally relevant degrees of freedom.
%%%%%%%%%%%%%%%%%%%%%%
%%%%%%%%%%%%%%%%%%%%%%%%%

\begin{figure}[t]
%[htb]
\includegraphics[width=1.0\linewidth]{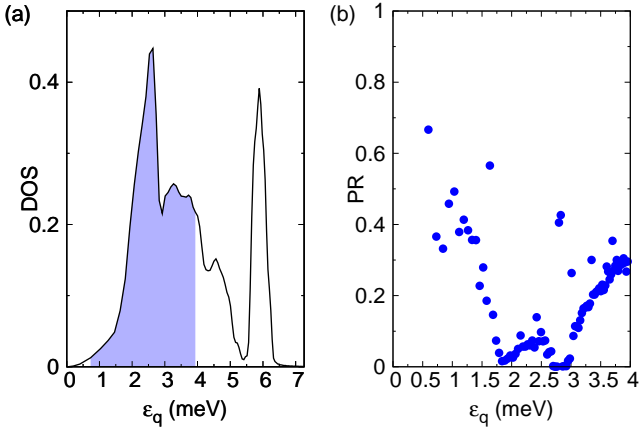}
\caption{(Color online) (a). Calculated phonon density of states (DOS) of off-center type-I BGS for the system size of 99$\times$99$\times$99 under periodic boundary condition. (b). Calculated participation ratio $P(\varepsilon_q)$ defined in Eq.\,(\ref{eq5}) as a function of eigenenergy $\varepsilon_q$ in the energy range marking by the blue shadow in (a)
for the system size of 20$\times$20$\times$20 under periodic boundary condition.
}
\label{fig33}
\end{figure}

\begin{figure}[t]
%[htb]
\includegraphics[width=0.9\linewidth]{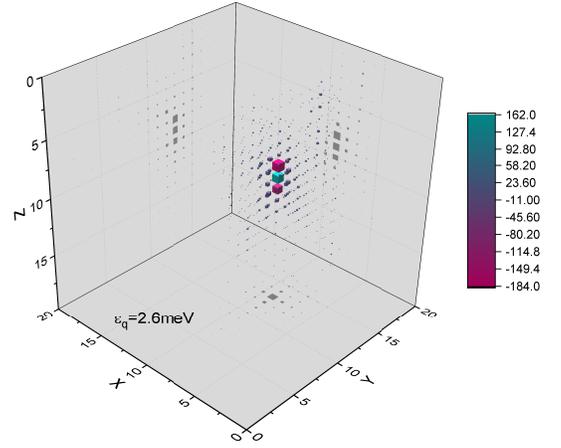}
\caption{(Color online) The mode pattern of SL modes belonging to the eigenenergy $\varepsilon_q$=2.6\,meV. Both the color scale and cubic size indicate the strength of amplitudes at each site. The mode pattern is obtained from the system size 20$\times$20$\times$20 under fixed boundary condition.}
\label{fig44}
\end{figure}

Extremely large system-sizes are required in computer simulations on disorder systems in order to distinguish localized modes from extended modes.
However, the present status of first-principles calculations (FPC) are limited to insufficient system-sizes for properly incorporating the disorder attributing to off-centeredness of guest atoms in  off-center type-I clathrates consisting of a unit cell with `54' atoms.
Thus, it is difficult not only to include realistic disorder reproducing glasslike thermal conductivities, but also to exclude finite size effect for propagating acoustic phonons.
Liu $et.~al.$~\cite{Liu2016} have performed calculations for 3D systems of (20$\times$20$\times$20)$\sim$(100$\times$100$\times$100)  molecular units, for which they have employed a powerful numerical method called the forced oscillator method.\cite{Williams1985, Nakayama2001}
They have also studied the localization nature of exited modes by taking the participation ratio (PR) as a criterion. The PR of a relevant mode \{$\varphi_\ell(\varepsilon_q); \ell=1,2,...N\}$ belonging to the eigenenergy $\varepsilon_q$ is defined by
\begin{eqnarray}
P(\varepsilon_q)=\frac{\left(\sum _{\ell=1}^N
\left|\varphi_\ell(\varepsilon_q)\right|^2\right)^2}{N\sum _{\ell=1}^N \left|\varphi_\ell(\varepsilon_q)\right|^4},
\label{eq5}
\end{eqnarray}
where $\ell$ denotes the $\ell$-th molecular unit depicted in Fig.\,$\ref{fig1}$\,(d) and $N$ is the total mode number. For EX modes in a finite system, $P(\varepsilon_q)$  take values close to $\approx$0.6 when $\varepsilon_q\neq0$, and $P(\varepsilon_q)$ becomes $\approx 1/N$ for SL modes \cite{Quasicrystals}. Figure~\ref{fig33}\,(a) is the calculated phonon density of states (DOS), and (b)\, the results of $P(\varepsilon_q)$ for the size of 20$\times$20$\times$20 lattice of off-center type-I BGS.
%We have omitted PRs below $\varepsilon_q$=0.5\,meV since those are obviously extended states of the Debye phonons as seen .
It is remarkable that $P(\varepsilon_q)$ ranges from a value of SL modes $P(\varepsilon_q)\approx 0$ to EX modes of $P(\varepsilon_q)\approx 0.6$.
We should emphasize that  there appear three kinds of modes in the THz frequency region and below classified into EX, WL and SL modes.
SL modes with PR values much smaller than unity are realized in the energy range from 2 to 3 meV as found from calculated mode patterns.
Figure \,\ref{fig44} depicts the mode patterns of
%WL at $\varepsilon_q$=1.7\,meV and
SL mode at $\varepsilon_q$=2.6\,meV.

 The calculations of the PR for excited modes depicted in Fig.\,\ref{fig33} have demonstrated that there exists  the delocalization-localization transition at a ``finite" frequency $\omega_c$ distinguishing EX and WL modes with the nature of acoustic modes vibrating ``in-phase" between guest ions and cages.
Furthermore, it has been found~\cite{Liu2016} that WL modes convert to SL modes at higher frquencies with the nature of optical modes vibrating ``out-of-phase" between guest ions and cages.
In this aspect, we note that Nakayama~\cite{Nakayama1998} had demonstrated the clear existence of the transition from WL to SL modes for the quasi-one-dimensional (1D) coarse-grained model consisting of host network and guest atoms connected by random springs.
It was found~\cite{Nakayama1998} that WL modes vibrate in-phase between network atoms and guest atoms, while SL modes manifest optical modes vibrating out-of-phase.
However, there is no  EX modes due to ``quasi-1D" model.
This manifests the Anderson weak localization criteria where the critical frequency $\omega_c$ takes a finite value in three dimensional (3D) systems, while it vanishes for 1D and 2D systems  suggesting no EX modes in 1D and 2D disordered systems.
The quasi-1D model~\cite{Nakayama1998} should be thought as the simplest theoretical model for cage-guest systems with broad implication for the dynamics of cage-guest systems.
%For example, the appearance of WL modes and Sl modes has been employed to interpret $\kappa_{\rm L}$ above the plateau in structural glasses\,\cite{Nakayama1999gg}.

%In this relevance, we note that Nakayama~\cite{Nakayama1998} had demonstrated the clear existence of the transition from WL to SL modes for the quasi-one-dimensional (1D) ladder model consisting of host network and guest atoms connected by random springs or masses, in which there is no EX modes according to 1D model.These are in accord with the case of the Anderson weak localization where the critical frequency $\omega_c$ takes a finite value in three dimensional (3D) systems, while it vanishes for 1D and 2D dimensional systems  suggesting no EX modes in 1D and 2D disordered systems.The ladder model~\cite{Nakayama1998} should be thought as the simplest model with deep implication for the dynamics of cage-guest systems.

%Figure~\ref{fig33}\,(b) depicts the eigenmode of SL mode belonging to the eigenfrequency $\varepsilon_q$=2.6\,meV, whose localization length $L(\varepsilon_q)$  is comparable with the wavelength $\lambda(\varepsilon_q)$  manifesting the Ioffe-Regel condition of SL modes.
%The SL modes are the hybridized modes between acoustic modes from network cages and local modes of guest ions in cages.

The observed delocalization-localization transition at $\varepsilon_q\thickapprox$\,1.3\,meV accords with the observed onset temperature of the plateau of $\kappa_{\rm L}$ in BGS at $T_{\rm P}\thickapprox$1.3\,meV/3.84$k_{\rm{B}}\thickapprox$\,3.9\,K as estimated from the Wien's displacement law for lattice thermal conductivities.
%This delocalization-localization transition is associated with the emergence of the plateau at around 4\,K in phonon thermal conductivities in BGS\,\cite{Avila2006} since the charcteristic temperature in phonon thermal conductivity corresponds to 1.5\,meV/3.84$\approx$4.5\,K from the Wien's displacement law.
Thus, the onset of the plateau is apparently due to the weak localization of acoustic modes.
The plateau region should be interpreted as the contribution of EX phonons ``saturates" at $T_{\rm P}$ for off-center type-I BGS. We note here that the random orientation of guest ions  in cages plays a crucial role to the localization.
%SL modes are fully excited above 3.0meV, which should correspond to the onset temperature of the subsequent T-linear increasing region $T_{\rm L}\thickapprox$3.0\,meV/3.84$k_{\rm{B}} \thickapprox$\,9\,K.
%We should remark that the onset temperature $T_{\rm P}$ is correlated with the strength of the coupling constant between guest ions and cages  generating the hybridization.

With increasing temperature further above a few 10K, $\kappa_{\rm L}$ show a linear rise on temperature\,\cite{Takabatake2014}.
This type of anomalous thermal conductivities characterized by the plateau and the subsequent {\it T}-linear rise of thermal conductivities have been clearly observed   for off-center type-I clathrates~\cite{Nolas:1998aa, Cohn:1999a, Sales:2001a, Paschen:2001a, Suekuni:2007a, Avila:2008aa, Suekuni:2008a}.
SL modes are fully excited above the temperature {\it T}$\simeq$10K$\approx$3\,meV/3.84$k_\mathrm{B}$ from the Wien's displacement law.
This indicates that {\it T}-linear rise subsequent to the plateau attributes to the excitations of SL modes.
In the following Sections, we present the theoretical interpretation on the underlying mechanism of  the linear rise on temperature above the plateau region for $\kappa_{\rm L}$.
%%%%%%%%%%%%%%%%%
%%%%%%%%%%%%%%%%%%%
%%%%%%%%%%%%%%%%
\section{Coarse-grained Hamiltonian for Type-I Off-center Clathrates}
%%%%%%%%%%%%%%%%%%%%%%%%%%%
\subsection{Harmonic Hamiltonian}
\begin{figure}[t]
%[htb]
\includegraphics[width=1.0\linewidth]{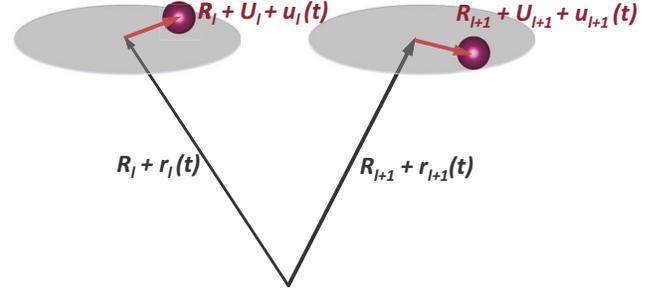}
\caption{(Color online) The definition of the position vectors:  $\RR_\ell+\rr_\ell(t)$ is the position vector of the $\ell$-th molecular unit at time $t$, where $\RR_\ell$ is the equilibrium position of the
$\ell$-th cage center and the vector $\rr_\ell(t)$ represents a small displacement from $\RR_\ell$ at time $t$.
The position vector of the guest ion(2) is defined by the vector $\RR_\ell+\UU_\ell+\uu_\ell(t)$, where $\UU_\ell$ is the equilibrium position of guest ion(2) from $\RR_\ell$, and $\uu_\ell(t)$ is a small displacement from $\RR_\ell+\UU_\ell$.
}
\label{fig22}
\end{figure}

%%%%%%%%%%%%%%%%%%%%%%%%%%
The Hamiltonian for off-center type-I clathrates under a coarse-grained picture consists of the kinetic energy of networked cages $K_{C}$ and off-center guest ions in cages $K_{G}$ in addition to the potential energy of the cage-cage interaction $V_{CC}$ and the cage-guest interaction $V_{CG}$.
This is expressed by
\begin{eqnarray}
H_0=K_{\textrm{C}}+K_{\textrm{G}}+V_{\textrm{CC}}+V_{\textrm{CG}}.
\end{eqnarray}
The explicit form of the total kinetic energy is given by the sum of $K_{C}$ and $K_{G}$ such as
\begin{eqnarray}
K=\frac{1}{2}\sum_{\ell}
\left(M\dot{\rr}_\ell(t)^2+m\dot{\uu}_\ell(t)^2\right),
\end{eqnarray}
where $m$  and $M$ are masses of the guest ion in tetrakaidecahedron cage and the remained molecular unit, respectively.
The vectors $\rr_\ell(t)$ and $\uu_\ell(t)$ represent small displacements of cage and guest ion from their equilibrium positions, $\RR_\ell$ and $\RR_\ell+\UU_\ell$, at the site $\ell$ as depicted in Fig.\,~\ref{fig22}.
Note here that guest ions take random orientation $\UU_\ell(\phi_\ell)$ in tetrakaidecahedron cages.

The molecular unit composed of tetrakaidecahedron and dodecahedron
is elastically connected with neighboring ones by the force constants $f_\|,f_\perp$.
These are related to the sound velocities of longitudinal ($\mu=\|$) and transverse ($\mu=\perp$) acoustic modes via the relation $v_\mu=a[f_\mu/(m+M)]^{1/2}$ with  $a=a_0/2$ where $a_0$ is the lattice spacing
of primitive cubic structure ($Pm\bar{3}n$) of type-I clathrates.
Thus, we can estimate the force constants $f_\|,f_\perp$ from the observed data of sound velocities.
Note here that 6 molecular units are included in unit cell in type-I clathrates.
In terms of these quantities, the potential energy of network cages becomes
\begin{eqnarray}
V_{CC}=\sum_{\ell'>\ell,\mu}
\frac{f_{\ell,\ell',\mu}}{2}
(\rr_{\ell,\mu}(t)-\rr_{\ell',\mu}(t))^2,
\label{eq22}
\end{eqnarray}
where $\mu=\|,\perp,\perp'$.
Hereafter, we keep up to the nearest neighbor coupling ($\ell'=\ell + 1$) between molecular units, which are denoted by $f_\|$, $f_\perp$ and $f_{\perp'}$.
% for the sake of simplicity.
%This type of coarse-grained description is effective to elucidate the underlying mechanism of glasslike $\kappa_{\rm L}(T)$ observed in type-I clathrates.
%We emphasize that the Hamiltonian Eq.\,(\ref{eq22}) describes the acoustic phonon propagating through periodically networked cages.
The effect of randomly  orientated guest ions are included in the following cage-guest interaction Hamiltonian.

The Hamiltonian  should satisfy the symmetry of infinitesimal translation-invariance as a whole, $i.e.$,
$\rr_\ell=\uu_\ell=\delta\aa$, which guarantees acoustic phonons  as the Nambu-Goldstone boson with the eigenfrequency $\omega_\kk\rightarrow 0$ for $\kk\rightarrow 0$.
This symmetry principle also holds for the potential of cage-guest interaction.
Hence, the potential function for the  cage-guest interaction $V_{CG}$ should be given by relative coordinates between the cage and the guest ion of
$\ww_\ell(t)=\uu_\ell(t)-\rr_\ell(t)$, which is
expressed by
%%%%%%%%%%%%%%%%%%%%%%%%%%%%%%%%%%%
\begin{eqnarray}
V_{CG}
=\sum_{\ell, m=\mathrm{in,out}}\frac{\xi_{m}}{2}~\ww_{\ell,m}^2(t),
\label{eq2x}
\end{eqnarray}
%%%%%%%%%%%%%%%%%%%%%%%%%%%%
where $\xi_m$ represents the force constants between cage and guest ion depending on in-plane (parallel) or out-of-plane motion (perpendicular) to the hexagonal face in the tetrakaidecahedron cage.
The guest ions execute in-plane vibration parallel to $x-y$ plane in addition to out-of-plane motions~\cite{Avila:2008aa} because of the anisotropic shape of tetrakaidecahedron cages.
This is because off-center guest ions are involved in tetrakaidecahedron cages whose shape distinguishes the vibrations of off-center guest ion(2) in the plane parallel and perpendicular to the hexagonal face of the cage.
Mori \textit{et~al}.\,\cite{Mori2011} observed  by means of THz time-domain spectroscopy that the lowest-lying peak of off-center BGS at 0.71\,THz splits into double peaks, $\omega_0^\phi/2\pi$=0.5THz and $\omega_0^r/2\pi$=0.72THz for off-center type-I BGS below T$\simeq$100\,K.
These spectra should be assigned to the libration and stretching modes of Ba(2) associated with $\xi_\phi$ and $\xi_r$.
The peak around 1.35\,THz is assigned as the out-of-plane motion of Ba(2) to the hexagonal faces of tetrakaidecahedron, which should be concerned with  $\xi_{\theta}$.
The Raman spectra of off-center $\rm Sr_{8}Ga_{16}Ge_{30}$ (SGG) have observed  A$_{1\mathrm{g}}$ stretching mode as 48\,cm$^{-1}$, and for off-center $\rm Eu_{8}Ga_{16}Ge_{30}$ (EGG) as 36\,cm$^{-1}$ at 2\,K\,\cite{Takasu2006}.
%Mori \textit{et~al}.\,\cite{Mori2011} observed  by means of THz time-domain spectroscopy that the lowest-lying peak of BGS at 0.71\,THz splits into double peaks, 0.5THz(2.07\,meV) and 0.72THz(2.98\,meV).This spectra should be assigned to the libration and stretching modes of Ba(2).The peak around 1.35\,THz(5.58\,meV) is assigned to $\xi_{\theta}$.
Using these data, we can estimate the force constants via the relation $\xi_{r,(\phi,\theta)}=m'\omega_{r,(\phi,\theta)}^2$, where $m'$ is the reduced mass defined by $1/m'=1/M+1/m$.

By taking account of this aspect, the quasi-harmonic Hamiltonian valid at {\it T}$\lesssim$100\,K, attributing to coupled vibrations between cages and guest atoms, can be expressed in the vector form as
\begin{eqnarray}
V_{CG}&=&\frac{1}{2}\sum_{\ell}\xi_r(\hat{\bm{U}}_\ell\cdot\bm{w}_{\parallel,_\ell})^2+\frac{1}{2}\sum_{\ell}\xi_{\phi}(\hat{\bm{U}}_\ell\times\bm{w}_{\parallel,_\ell})^2\nonumber\\
&+&\frac{1}{2}\sum_{\ell}\xi_{\theta}(\bm{w}{\perp,_\ell})^2,
\label{eq88vec}
\end{eqnarray}
where $\hat{\bm{U}}_\ell=(\hat{U}_\ell^x, \hat{U}_\ell^y)$
is the unit vector for the vector $\bm{U}_\ell$.
$\lbrace\phi_\ell\rbrace$ and $\lbrace\theta_\ell\rbrace$ represent the azimuthal  and the polar angle in spherical coordinates.
The effect of ``random" orientation of guest ions $\lbrace\phi_\ell\rbrace$ induced by off-centeredness are involved in $\left\lbrace\UU_\ell\right\rbrace$.
The relation between off-centeredness and disorder in Eq.\,(\ref{eq88vec}) is described in details in Supplemental Material (SM).

%%%%%%%%%%%
%%%%%%%%%%%
\subsection{Anharmonic coupling between acoustic phonons and SL modes}
%%%%%%%%%%%%%%%%%%%%%%%%%%%%
%Raman scattering experiments for off-center type I BGS~\cite{Kume2015} have provided the evidence that low-lying in-plane modes at $\nu\sim$0.5\,THz are sensitive to pressure. These data indicate that the anharmonic effect between become relevant in the 1st and 2nd terms in Eq.\,(\ref{eq88vec}).
%Spectroscopic data for off-center type-I clathrates~\cite{ Takasu2006, Mori2011} have provided the evidence that low-lying in-plane modes at $\nu\sim$0.5\,THz relevant to guest ions depend strongly  on temperature, while the modes belonging to higher eigenfrequencies $\nu\gtrsim$1\,THz related to the vibrations of out-of-plane motions and of cage themselves do weakly on temperature. Hence, it is reasonable to take into account the anharmonic effect between acoustic mode and  in-plane modes for the 1st and 2nd terms in Eq.\,(\ref{eq88vec}).
%\blue{in comparison with the third term}.
When acoustic modes (LA and TA) are propagating along networked cages, the cages are distorted and these change the states of guest ions, which are realized via the change of the force constants  $\xi_r$ and $\xi_\phi$ in Eq.\,(\ref{eq88vec}).
The in-plane (stretching and libration) modes are sensitive to temperature/pressure compared with out-of-plane modes as shown in the optic spectroscopy data below {\it T}$\simeq$100\,K .\cite{Mori2011, Takasu2006}
Thus, the anharmonic effect between  acoustic modes and in-plane modes in the first and the second terms in Eq. (\ref{eq88vec}) becomes  relevant in comparison with the third term.
The expansions of $\xi_r$ and $\xi_\phi$ with respect to the strain tensor $e_{\alpha\beta}$ for $\alpha,\beta=x,y,z$  provide
\begin{eqnarray}
\xi_{r,(\phi)}&=&\xi{_{r,(\phi)}^{(0)}}+\sum_{\alpha=x,y,z}
%\frac{\partial \xi_r}{\partial e_{\alpha\alpha} }
D_{r,(\phi)}e_{\alpha\alpha}+\sum_{\alpha,\beta=x,y,z \atop \alpha\neq\beta}
%\frac{\partial \xi_{\theta}}{\partial e_{\alpha\beta}}
S_{r,(\phi)}e_{\alpha\beta}+...,
%\nonumber\\
%&+&\frac{1}{2}\sum_{\substack{\alpha=x,y,z\\ \alpha'=x,y,z}}\frac{\partial^2 \xi_{r}}{\partial e_{\alpha\alpha}\partial e_{\alpha'\alpha'}}e_{\alpha\alpha}e_{\alpha'\alpha'}+...,
\label{eq2anh}
\end{eqnarray}
%and
%\begin{eqnarray}
%\xi_{\phi}&=&\xi{_{\phi}^{(0)}}+\sum_{\alpha=x,y,z}D_\phi e_{\alpha\alpha}+\sum_{\alpha,\beta=x,y,z \atop \alpha\neq\beta}
%\frac{\partial \xi_{\theta}}{\partial e_{\alpha\beta}}S_\phi e_{\alpha\beta}+...,
%\nonumber\\
%&+&\frac{1}{2}\sum_{\substack{\alpha,\beta=x,y,z\\ \alpha',\beta'=x,y,z}}\frac{\partial^2\xi_\phi}{\partial e_{\alpha \beta}\partial e_{\alpha' \beta'}}e_{\alpha\beta}e_{\alpha'\beta'}+....
%\label{eq3anh}\end{eqnarray}
Here the coefficients are defined by $D_{r,(\phi)}=\partial \xi_{r,(\phi)}/\partial e_{\alpha\alpha}$,  %$D_\phi=\partial \xi_\phi/\partial e_{\alpha\alpha}$,
$S_{r,(\phi)}=\partial \xi_{r,(\phi)}/\partial e_{\alpha\beta(\alpha\neq\beta)}$
% and $S_\phi=\partial \xi_{\phi}/\partial e_{\alpha\beta(\alpha\neq\beta)}$,
where
$e_{\alpha\beta}=1/2\left(\partial u_\alpha/\partial x_\beta+\partial u_\beta/\partial x_\alpha\right)$ is the component of strain tensor.
It should  be noted that $e_{\alpha\alpha}$ expresses the compression or expansion, and $e_{\alpha\beta(\alpha\neq\beta)}$ does the shear destorsion.
%\red{\begin{equation}
%{CG}=\frac{1}{2}\sum_{\ell,\alpha} \underbrace{\frac{\partial \xi_r}{\partial e_{\alpha\alpha}}}_{=\,D}e_{\alpha\alpha}(\delta r_\ell)^2+\frac{1}{2}\sum_{\ell,\alpha\neq\beta}\underbrace{\frac{\partial \xi_{\theta}}{\partial e_{\alpha\beta}}}_{=\,\red{S}}e_{\alpha\beta}(r_0\delta \phi_\ell)^2
%\nonumber\\
%&+&\frac{1}{2}\sum_{\ell,\alpha}\underbrace{\frac{\partial \xi_z}{\partial e_{\alpha\alpha}}}_{=\,Z}e_{\alpha\alpha}(U_0\delta \phi_\ell)^2.
%\label{eq456}\end{equation}}
%and the second order of the form
%\begin{equation}\begin{split}
%V''_{CG}=\frac{1}{4}\sum_{\ell,\alpha,\alpha'}\frac{\partial^2 \xi_{r}}{\partial e_{\alpha\alpha}\partial e_{\alpha'\alpha'}}e_{\alpha\alpha}e_{\alpha'\alpha'}(\delta r_{l})^2+\frac{1}{4}\sum_{\substack{\ell,\alpha\neq\beta, \\ \alpha'\neq\beta'}}\frac{\partial^2\xi_\theta}{\partial e_{\alpha \beta}\partial e_{\alpha' \beta'}}e_{\alpha\beta}e_{\alpha'\beta'}(r_0\delta \theta_\ell)^2 \\+\frac{1}{4}\sum_{\ell,\alpha,\alpha'}\frac{\partial^2 \xi_z}{\partial e_{\alpha\alpha}\partial e_{\alpha'\alpha'}}e_{\alpha\alpha}e_{\alpha'\alpha'}(U_0\delta \phi_\ell)^2\end{split}\label{eq44}\end{equation}
The expansion in Eq.\,(\ref{eq2anh}) leads to the following anharmonic interaction expressed in the vector form as
\begin{equation}
\begin{split}
V'_{CG}=\frac{1}{2}\sum_{\ell,\alpha\neq\beta}(D_r e_{\alpha\alpha}+S_re_{\alpha\beta})(\hat{\bm{U}}_\ell\cdot\bm{w}_{\parallel,\ell})^2
\\
+\frac{1}{2}\sum_{\ell,\alpha\neq\beta}(D_\phi e_{\alpha\alpha}+S_\phi e_{\alpha\beta})(\hat{\bm{U}}_\ell\times\bm{w}_{\parallel,\ell})^2.
\end{split}
\label{eq88}
\end{equation}
Here we note that Eq. (\ref{eq88}) satisfies the condition of infinitesimal translational invariance as a whole; $V'_{CG}\rightarrow 0$ under the long wavelength limit $k_{\mu}\rightarrow 0$.
We emphasize again that Eq.\,(\ref{eq88}) is valid at temperatures {\it T}$\lesssim$100\,K where the guest atoms execute coupled vibrations with cages.\cite{Mori2011, Takasu2006}
While, at {\it T}$\gtrsim$100\,K, $\kappa_L$({\it T}) saturates without exhibiting the appreciable {\it T}-dependence, where guest atoms behave like rattlers in cages termed by the "rattling" motion, where the concept of vibrational modes is invalid.\cite{Mori2011, Takasu2006}

%%%%%%%%%%%%%%%%%%
%%%%%%%%%%%%%%%%%%%%%%
\section{The 2nd quantized form of interaction Hamiltonian}
\subsection{Acoustic phonons causing from networked cages}
Provided that EX acoustic phonons with wavelengths $\lambda$ much larger than the lattice spacing $a_0$ propagate through networked cages, the molecular units  and guest ions vibrate ``in phase".
%These modes  \green{follow} the linear dispersion relation of  $\omega_\mu=v_\mu k_\mu$. For EX acoustic modes,
The displacement at the site $\ell$ is expressed by the sum of plane waves as given by
\begin{eqnarray}
\rr_{\ell}(t)=\sum_{\kk_\mu}\sqrt{\frac{\hbar}{2{\rho}\omega_{\kk_\mu}}}\hat{\bm{e}}_{\kk_\mu}\left(\phi_{\kk_\mu}(\RR_\ell)b{_{\kk_\mu}^\dagger}(t)+h.c.\right).
\label{eq9ex}
\end{eqnarray}
Here the symbols $b{_{\kk_\mu}^\dagger}\,(b_{\kk_\mu})$ express the creation (annihilation) operator for acoustic phonon of the mode $(\kk_\mu)$ with $\mu=\parallel, \perp$, which represent longitudinal and transverse modes, respectively.
The vector $\bm{R}_\ell$ expresses  the equilibrium
position of the $\ell$th molecular unit as depicted in Fig.\,\ref{fig22}, and $h.\,c.$\,indicates the Hermitian conjugate.
The mass density is defined as $\rho=6(m+M)/a_0^3$ with the size of unit cell of $a_0$ since 6 molecuar units are involved in unit cell of type-I clathrates.
See Sec. I in Supplemental Material (SM) about the definitions employed in this paper.

The function $\phi_{\kk_\mu}(\RR_\ell)$ in Eq.\,(\ref{eq9ex}) takes the form of
\begin{equation}
\phi_{\kk_\mu}(\RR_\ell)= \sqrt{\frac{1}{V}}e^{i\kk_\mu\cdot\RR_\ell}.
\label{eq99}
\end{equation}
The normalization condition for $\phi_{\kk_\mu}(\RR_\ell)$ is given by
\begin{equation}
\int\mid\phi_{\kk_\mu}(\RR_\ell)\mid^2d\RR_\ell=1 .
\label{eq999}
\end{equation}

%Note here that Eq.~(\ref{eq9}) takes also the form
%\begin{equation}\rr_{\ell}(t)=\sum_{\kk_\mu}\sqrt{\frac{\hbar V}{2N(m+M)\omega_{\kk_\mu}}}\hat{\bm{e}}_{\kk_\mu}\left(\phi_{\kk_\mu}(\RR_\ell)b{_{\kk_\mu}^\dagger}(t)+h.c.\right);\rho=\frac{N(m+M)}{V}=\frac{m+M}{a^3}.\label{eq9911}\end{equation}

%%%%%%%
\subsection{Strongly localized modes due to guest ions}

%Randomly oriented off-center guest ions are weakly connected to tetrakaidecahedron cages.
%, and those force constants take orientation-dependent random values  leading the localization of relevant vibrational modes.
Figure \ref{fig44} provides the mode belonging to the eigenenergy $\varepsilon_q$=2.6\,meV obtained for the system size 20$\times$20$\times$20.
% under fixed boundary condition.
This mode pattern indicates
 that the localization length $L_{\lambda}$ is comparable with the wavelength $2\pi/k_\lambda$, $i. e.$, localized within several molecular units, manifesting the Ioffe-Regel condition of the strong localization.
%Since the mode pattern in Fig.\,\ref{fig44} is depicted in terms of the relative coordinates $\ww_\ell(t)=\rr_\ell(t)-\uu_\ell(t)$ defined in Eq.\,(\ref{eq2x}), the mode possesses the nature of ``optic" modes vibrating in ``out-of-phase" between the cages and guest ions.This accords with SL modes in quasi-1D model found in Ref.~\cite{Nakayama1998}.
On the basis of these numerical findings, we can express the form of SL modes
in terms of the relative coordinate $\ww_\ell(t)=\uu_l(t)-\rr_{l}(t)$ as
\begin{equation}
\ww_\ell(t)= \sum_{\lambda}\sqrt{\frac{\hbar}{2m'\omega_{\lambda}}}\hat{\bm{e}}
_{\lambda}\left(\psi_{\lambda}(\RR_\ell)c_{\lambda}^\dagger(t)+h.c.\right).
%= \sum_{\lambda}\sqrt{\frac{\hbar}{2\rho'\omega_{\lambda}}}\hat{\bm{e}};~~~\rho'm'/a63=Nm'/V_{\lambda}.
\label{eq9999}
\end{equation}
Here the mass $m'$ is the reduced mass defined by $1/m'=1/M+1/m$, where
$M$ is the mass of the molecular unit given in Fig.\,\ref{fig1}, much larger than the mass of guest ion $m$, for example, $M=6.01m$ for off-center type-I BGS.
The symbol $c{_{\lambda}^\dagger}\,(c_{\lambda})$ represents the creation (annihilation) operator for the localized mode $\lambda$.
We put forward the Ansatz for the amplitude $\psi_{\lambda}(\RR_\ell)$ of the form
\begin{equation}
\psi_{\lambda}(\RR_\ell)= A\cos\left[ \kk_{\lambda}\cdot(\RR_\ell-\RR_{\lambda})\right] e^{-|\RR_\ell-\RR_\lambda|/L_\lambda}.
\label{eq91}
\end{equation}
where $\RR_\lambda$ represents the center of SL mode $\lambda$.
This wave function  has vanishing group-velocities $v_g$ characterizing localized modes.

%The Ioffe-Regel condition is expressed by the form $k_\lambda^r L_\lambda\approx 2\pi$ for the localization length $L_\lambda$,  so $k_\lambda^r$ is given by $k_\lambda^r\approx 2\pi/L_\lambda$.

The prefactor $A$ in Eq.\,(\ref{eq91}) can be determined from the normalization condition of
\begin{equation}
\sum_\ell\mid\psi_{\lambda}(\RR_\ell)\mid^2=\frac{1}{\Omega}\int d\RR_\ell\mid\psi_{\lambda}(\RR_\ell)\mid^2=1,
\label{eq92}
\end{equation}
where $\Omega=V/N$ is the volume of the molecular unit depicted in Fig.\,\ref{fig1}(d).
This yields, by combing with the Ioffe-Regel condition,
\begin{equation}
A\cong\sqrt{\frac{2\Omega}{\pi L_\lambda^3}}.
\label{eq9222AA}
\end{equation}
The above has been obtained by using the formula $\cos^2(\kk\cdot\RR)=(\cos(2\kk\cdot\RR)+1)/2$.
According to the Ioffe-Regel condition $k\approx2\pi/L_\lambda$, the 1st term in the integral becomes negligible compared with the 2nd term  since the 1st term  yields rapidly oscillating function in the integrand.
This leads to Eq.\,(\ref{eq9222AA}).
Thus, the normalized wave function of the SL mode
$\lambda$ becomes
\begin{equation}
\psi_{\lambda}(\RR_\ell)= \sqrt{\frac{2\Omega}{\pi L_\lambda^3}}\cos\left[ \kk_{\lambda}\cdot(\RR_\ell-\RR_{\lambda})\right] e^{-|\RR_\ell-\RR_\lambda|/L_\lambda}.
\label{eq93}
\end{equation}

%\red{We can also use the relation $a^3/m'=1/\rho'$, which makes the expression short, but we proceed by using $m'$}.\\

%%%%%%%%%%%%%%%%%%%%%%%%%%%%%%%%%%%%%%
\subsection{Anharmonic Hamiltonian between SL and EX modes}
%%%%%%%%%%%%%%%%%%
%%%%%%%%%%%%%%%%%%%%%
We consider here the effect of incoming EX acoustic phonons with the polarization vector $\hat{\ee}_{\kk_\mu}$ to SL modes with the polarization vectors  $\hat{\ee}_{\lambda'}$ and $\hat{\ee}_{\lambda''}$.
These are included in Eq.\,(\ref{eq88}) as the scalar product $(\hat{\ee}_{\lambda'}\cdot\hat{\UU}_\ell)(\hat{\ee}_{\lambda''}\cdot\hat{\UU}_\ell)$ and the  product $(\hat{\ee}_{\lambda'}\times\hat{\UU}_\ell)\cdot(\hat{\ee}_{\lambda''}\times\hat{\UU}_\ell)$.
At first, we fix the direction of the wave vector of incoming EX phonons $\kk_{\mu}$ and later we include the contributions from 3 components of the wave vector $\kk_{\mu}$.
%While, only two polarizations of SL mode are involved in the anharmonic interaction since $\ww_{\parallel,\ell}$ is constrained in plane.
We should note that the deformation (normal or shear strain) of cages causing from incoming acoustic phonons responses to every directions of the polarization vector of SL modes,
which provides both the interaction  between the same polarization and different polarizations of SL modes as shown below.
%the extra factor of 2$\times$2=4 from the product $(\hat{\ee}_{\lambda'}\cdot\hat{\UU}_l)( \hat{\ee}_{\lambda''}\cdot\hat{\UU}_l)$ and the  product $(\hat{\ee}_{\lambda'}\times\hat{\UU}_\ell)\cdot(\hat{\ee}_{\lambda''}\times\hat{\UU}_\ell)$ as shown below.
%%%%%%%%%%%%%%%%%
\begin{figure}[t]
%[htb]
\includegraphics[width=1.0\linewidth]{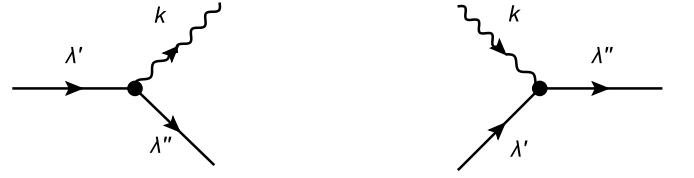}
\caption{(Color online) The diagrams showing the hopping process for strongly SL modes arising from anharmonic interaction between SL modes and EX modes: (a) SL $\rightarrow$ EX + SL, and (b) EX + SL $\rightarrow$ SL. The solid lines denote SL mode and the wavy lines EX mode. }
\label{fig55}
\end{figure}
%%%%%%%%%%%%%%%%

The second quantized anharmonic Hamiltonian is obtained  by substituting  Eqs.\,(\ref{eq9ex}) and (\ref{eq9999}) into Eq.\,(\ref{eq88}) by using the relations
given in Sec.\,II in SM.
The product of the field operators $b_{\kk_\mu} c_{\lambda'} c_{\lambda''}$ consists of eight terms.  The two involve the combinations  $b_{\kk_\mu}^\dag c_{\lambda'}^\dag c_{\lambda''}^\dag$ and $b_{\kk_\mu}c_{\lambda'}c_{\lambda''}$ are irrelevant to the hooping processes because of not conserving the total energy.
Furthermore the other two terms $b_{\kk_\mu}^\dag c_{\lambda'}c_{\lambda''}$ and $b_{\kk_\mu}c_{\lambda'}^\dag c_{\lambda''}^\dag$ do not contribute
to the scattering processes since the energies of EX modes are smaller than those of SL modes.
Hence, the relevant second quantized anharmonic Hamiltonian for the process on EX + SL $\rightarrow$ SL is given by
\begin{equation}
\begin{split}
H_{CG}'
&=\sum_{\kk_\mu,\lambda',\lambda''}(A_{\kk_\mu,\lambda',\lambda''}
b_{\bm{k}_\mu}c_{\lambda'}c{_{\lambda''}^\dagger}+h.c.),
\\
&+\sum_{\kk_\mu,\lambda''',\lambda''''}(B_{\bm{k}_\mu,\lambda''',\lambda''''}
b_{\bm{k}_\mu}c_{\lambda'''}c{_{\lambda''''}^\dagger}+h.c.)
\\
&+\sum_{\kk_\mu,\lambda',\lambda'''}(C_{\bm{k}_\mu,\lambda',\lambda'''}
b_{\bm{k}_\mu}c_{\lambda'}c{_{\lambda'''}^\dagger}+h.c.),
\label{eqABC}
\end{split}
\end{equation}
where $A_{\kk_\mu,\lambda',\lambda''}$ is associated with the interaction between
the modes with $x$-polarization, $B_{\kk_\mu,\lambda''',\lambda''''}$ corresponds to the interaction between $y$-polarization, and $C_{\kk_\mu,\lambda',\lambda'''}$ does the interaction between two different polarizations.
See Fig.\,\ref{fig55}.
%The explicit expressions on these quantities are given in Sec.\,III in SM.

By taking the unit vectors $\hat{x},\hat{y},\hat{z}$  the same as the directions of the polarizations $\hat{\ee}_\parallel,\hat{\ee}_{\perp},\hat{\ee}_{\perp'}$ of EX acoustic modes, we have
\begin{equation}
\begin{split}
&A_{\kk_\mu,\lambda',\lambda''}
=-\frac{1}{4}\sum_{l}i\sqrt{\frac{\hbar}{2\rho\omega_{\kk_\mu}}}\sqrt{\frac{\hbar}{2m'\omega_{\lambda'}}}\sqrt{\frac{\hbar}{2m'\omega_{\lambda''}}}
\\
&\times\phi_{\kk_\mu}\psi_{\lambda'}\psi_{\lambda''}
\left[ (D_r+D_\phi)k_{\parallel}\delta_{\mu,\parallel}+(S_r+S_\phi)k_{\perp}\delta_{\mu,\perp}\right],
\end{split}
\label{eq33A}
\end{equation}
and the term on $B_{\kk_\mu,\lambda''',\lambda''''}
$ becomes the same as $A_{\kk_\mu,\lambda',\lambda''}$ by setting [$\lambda', \lambda''\longrightarrow \lambda''', \lambda''''$].
%\begin{equation}\begin{split}
%&B_{\kk_\mu,\lambda''',\lambda''''}=-\frac{1}{4}\sum_{l}i\sqrt{\frac{\hbar}{2\rho\omega_{\kk_\mu}}}\sqrt{\frac{\hbar}{2m'\omega_{\lambda'''}}}\sqrt{\frac{\hbar}{2m'\omega_{\lambda''''}}}
%\\&\times\phi_{\kk_\mu}\psi_{\lambda'''}\psi_{\lambda''''}
%\left[ (D_r+D_\phi))k_{\parallel}\delta_{\mu,\parallel}+(S_r+S_\phi)k_{\perp}\delta_{\mu,\perp}\right],
%\end{split}\label{eq33B}\end{equation}
The last one should be
\begin{equation}
\begin{split}
&C_{\kk_\mu,\lambda',\lambda'''}
=-\frac{1}{\pi}\sum_{l}i\sqrt{\frac{\hbar}{2\rho\omega_{\kk_\mu}}}\sqrt{\frac{\hbar}{2m'\omega_{\lambda'}}}\sqrt{\frac{\hbar}{2m'\omega_{\lambda'''}}}
\\
&\times\phi_{\kk_\mu}\psi_{\lambda'}\psi_{\lambda'''}
\left[(D_r-D_\phi)k_{\parallel}\delta_{\mu,\parallel}+(S_r-S_\phi)k_{\perp}\delta_{\mu,\perp}\right].
\end{split}
\label{eq33C}
\end{equation}
%From these we can obtain the squared quantities of the above relations.
The squared quantity on Eq.\,(\ref{eq33A}) is given by
\begin{equation}
\begin{split}
A^2_{\bm{k}_\mu,\lambda',\lambda''}
=&\frac{CI_1^2}{VL_{\lambda'}^3L_{\lambda''}^3}\frac{1}{\omega_{\kk_\mu}\omega_{\lambda'}\omega_{\lambda''}}
\\
&\left[ (D_r+D_\phi)k_{\parallel}\delta_{\mu,\parallel}+(S_r+S_\phi)k_{\perp}\delta_{\mu,\perp}\right]^2,
\end{split}
\label{eq50}
\end{equation}
where the coefficient $C$ is defined as
\begin{equation}
C=\frac{\hbar^3\Omega^2}{2^5\rho m'^2}.
\label{eq51}
\end{equation}
The expression of $B_{\kk_\mu,\lambda''',\lambda''''}^2$ takes the same form as $A^2_{\bm{k}_\mu,\lambda',\lambda''}$, since they both correspond to the interaction between SL modes with the same polarization. While $C_{\kk_\mu,\lambda',\lambda'''}^2$ corresponding to interaction between different polarizations has an additional factor $(4/\pi)^2$ and $\left[ (D_r-D_\phi)k_{\parallel}\delta_{\mu,\parallel}+(S_r-S_\phi)k_{\perp}\delta_{\mu,\perp}\right]^2$.
% instead of $\left[ (D_r+D_\phi)k_{\parallel}\delta_{\mu,\parallel}+(S_r+S_\phi)k_{\perp}\delta_{\mu,\perp}\right]^2$.
%%%%%%%%%%%%%%%%%%%%%%
%%%%%%%%%%%%%%%%%%%%%%%
\section{Hopping Process}
%%%%%%%%%%%%%%%%%%%%%%%%%%%%
%\section{ The decay of SL modes due to anharmonic interaction}
\subsection{Relaxation time of SL modes}
This subsection gives the formula for the relaxation time of SL mode due to the scattering process EX+SL$\rightarrow$ SL (hopping process) together with its reverse process shown in Fig.\,\ref{fig55} by applying the Fermi golden rule.
%This provides the transition probability $W_{\kk_\mu,\lambda'\to\lambda''}=1/\tau_{\lambda'}$ for the above process of the form,
To obtain the total transition rate of the SL mode in $\lambda'$, we have to incorporate all of four processes for each polarization as given below.
%\red{The following form is inconsistent  because the left is for the decay of the mode $\lambda'$, but the rights include $\lambda''', \lambda''''$ from the second line. Please write down the proper expression.}
%%%%%%%%%%%%%%%%%%%%%%%%%
These provide the decay of the Bose-Einstein distribution function $n_{\lambda'}$ for the occupied state $\lambda'$,
\begin{eqnarray}
\frac{dn_{\lambda'}}{dt}
&=&
\frac{2\pi}{\hbar^2}\sum_{\kk_\mu,\lambda''}|A_{\kk_\mu,\lambda',\lambda''}|^2[n_{\lambda''}(1+n_{\kk_\mu})(1+n_{\lambda'})
\nonumber\\
&-&n_{\kk_\mu}n_{\lambda'}(1+n_{\lambda''})]\delta(\omega_{\lambda''}-\omega_{\lambda'}-\omega_{\kk_\mu})
\nonumber\\
&+&|A_{\kk_\mu,\lambda'',\lambda'}|^2
[n_{\kk_\mu}n_{\lambda''}(1+n_{\lambda'})-n_{\lambda'}(1+n_{\kk_\mu})
\nonumber\\
&\times&(1+n_{\lambda''})]\delta(\omega_{\lambda'}-\omega_{\lambda''}
-\omega_{\kk_\mu})
\nonumber\\
&+&\left[A_{\kk_\mu,\lambda',\lambda''}\longrightarrow C_{\kk_\mu,\lambda',\lambda'''},\lambda''\longrightarrow\lambda'''\right]
% + \left[ C_{\kk_\mu,\lambda',\lambda'''}\right]
%&+&\frac{2\pi}{\hbar^2}\sum_{\kk_\mu,\lambda'''}|C_{\kk_\mu,\lambda',\lambda'''}|^2[n_{\lambda'''}(1+n_{\kk_\mu})(1+n_{\lambda'})
%\nonumber\\
%&-&n_{\kk_\mu}n_{\lambda'}(1+n_{\lambda''})]\delta(\omega_{\lambda'''}-\omega_{\lambda'}-\omega_{\kk_\mu})
%\nonumber\\
%&+&|C_{\kk_\mu,\lambda''',\lambda'}|^2[n_{\kk_\mu}n_{\lambda'''}(1+n_{\lambda'})-n_{\lambda'}(1+n_{\kk_\mu})
%\nonumber\\&\times&(1+n_{\lambda'''})]\delta(\omega_{\lambda'}-\omega_{\lambda'''}-\omega_{\kk_\mu}) .
%&+&|B_{\kk_\mu,\lambda'''',\lambda'''}|^2[n_{\kk_\mu}n_{\lambda''''}(1+n_{\lambda'''})-n_{\lambda'''}(1+n_{\kk_\mu})(1+n_{\lambda''''})]\delta(\omega_{\lambda'''}-\omega_{\lambda''''}-\omega_{\kk_\mu})&+&\frac{2\pi}{\hbar^2}\sum_{\kk_\mu,\lambda''}|C_{\kk_\mu,\lambda',\lambda'''}|^2[n_{\lambda'''}(1+n_{\kk_\mu})(1+n_{\lambda'})-n_{\kk_\mu}n_{\lambda'}(1+n_{\lambda'''})]\delta(\omega_{\lambda'''}-\omega_{\lambda'}-\omega_{\kk_\mu})\nonumber\\&+&|C_{\kk_\mu,\lambda''',\lambda'}|^2[n_{\kk_\mu}n_{\lambda'''}(1+n_{\lambda'})-n_{\lambda'}(1+n_{\kk_\mu})(1+n_{\lambda'''})]\delta(\omega_{\lambda'}-\omega_{\lambda'''}-\omega_{\kk_\mu})
\label{eq40REX}
\end{eqnarray}
%%%%%%%
%%%%%%%%%%%%%
We consider, at first, the decay due to the hopping process between the same polarization, $i.e.$, the contribution from the first two terms of Eq.\,(\ref{eq40REX}).
By separating the distribution function into two parts. $n=n^{(0)}+n^{(1)}$, where $n^{(0)}$ is the Bose-Einstein distribution function in equilibrium state and $n^{(1)}$ is its deviation due to the scattering processes,
and
%%%%%%%%
%We use the relation of the form
%\begin{equation}
%\begin{split}&(n{_{\lambda''}^{(0)}}+n{_{\lambda''}^{(1)}})(1+n{_{\kk\mu}^{(0)}}+n{_{\kk\mu}^{(1)}})(1+n{_{\lambda'}^{(0)}}+n{_{\lambda'}^{(1)}})\\&-(n{_{\kk\mu}^{(0)}}+n{_{\kk\mu}^{(1)}})(n{_{\lambda'}^{(0)}}+n{_{\lambda'}^{(1)}})(1+n{_{\lambda''}^{(0)}}+n{_{\lambda''}^{(1)}})\\&=(n{_{\lambda''}^{(0)}}-n{_{\lambda'}^{(0)}})n{_{\kk\mu}^{(1)}}+(1+n{_{\kk\mu}^{(0)}}+n{_{\lambda'}^{(0)}})n{_{\lambda''}^{(1)}}
%\\&+(n{_{\lambda''}^{(0)}}-n{_{\kk\mu}^{(0)}})n{_{\lambda'}^{(1)}}+n{_{\lambda''}^{(1)}}n{_{\kk\mu}^{(1)}}+n{_{\lambda''}^{(1)}}n{_{\lambda'}^{(1)}}
%-n{_{\lambda'}^{(1)}}n{_{\kk\mu}^{(1)}},
%\end{split}\end{equation}
%where the following identity is used under the energy conservation condition $\omega_{\lambda''}=\omega_{\lambda'}+\omega_{\kk\mu}$,
%\begin{equation}
%n{_{\lambda''}^{(0)}}(1+n{_{\kk\mu}^{(0)}})(1+n{_{\lambda'}^{(0)}})-n{_{\kk\mu}^{(0)}}n{_{\lambda'}^{(0)}}(1+n{_{\lambda''}^{(0)}})=0.
%\end{equation}
by employing the relaxation time approximation, $dn{_{\lambda'}^{(1)}}/dt=-n{_{\lambda'}}^{(1)}/\tau_{\lambda'}$, we have the inverse of relaxation time from Eq.\,(\ref{eq40REX})
for the same polarization process,
%\begin{eqnarray}
\begin{eqnarray}
%\begin{equation}\begin{split}
\frac{1}{\tau_{\lambda'}^{\rm same}}
&\cong&
%\frac{2\pi}{\hbar^2}\sum_{(\kk,\mu),\lambda''}|&A_{\bm{k},\mu,\lambda',\lambda''}|^2\delta(\omega_{\lambda''}-\omega_{\lambda'}-\omega_{\kk\mu})(n{_{\kk\mu}^{(0)}}-n{_{\lambda''}^{(0)}})\\+&|A_{\bm{k},\mu,\lambda'',\lambda'}|^2\delta(\omega_{\lambda'}-\omega_{\lambda''}-\omega_{\kk\mu})(1+n{_{\kk\mu}^{(0)}}+n{_{\lambda''}^{(0)}})\\=
\frac{2\pi}{\hbar^2}
\frac{CI_1^2}{VL^6}
\sum_{\kk_\mu,\lambda''}
\frac{1}{\omega_{\kk\mu}\omega_{\lambda'}\omega_{\lambda''}}
\nonumber\\
&\times&\left[ (D_r+D_\phi)k_{\parallel}\delta_{\mu,\parallel}+(S_r+S_\phi)k_{\perp}\delta_{\mu,\perp}\right]^2
\nonumber\\
&\times&
[\delta(\omega_{\lambda''}-\omega_{\lambda'}-\omega_{\kk\mu})(n{_{\kk\mu}^{(0)}}-n{_{\lambda''}^{(0)}})
\nonumber\\
&+& \delta(\omega_{\lambda'}-\omega_{\lambda''}-\omega_{\kk\mu})(1+n{_{\kk\mu}^{(0)}}+n{_{\lambda''}^{(0)}})],
%\end{split}
\label{eq46}
\end{eqnarray}
where the explicit form of the summation $I_1$ arising from the overlapping of wave functions $\psi_{\lambda'}$ and  $\psi_{\lambda''}$ is given by
\begin{eqnarray}
I_1&=&\sum_{\ell}e^{-i\kk_\mu\cdot\RR_l}\cos\left[ \kk_{\lambda'}\cdot(\RR_l-\RR_{\lambda'})\right]
e^{-|\RR_\ell-\RR_{\lambda'}|/L_{\lambda'}}
\nonumber\\
&\times&
\cos\left[ \kk_{\lambda''}\cdot(\RR_l-\RR_{\lambda''})\right] e^{-|\RR_\ell-\RR_{\lambda''}|/L_{\lambda''}}.
\label{eq32I1}
\end{eqnarray}
The above sum $I_1$ can be reduced to, by taking the origin of the sum as $\RR_{\lambda'}=0$ and the nearest neighbor position from the origin as $\RR_{\lambda''}=\Delta\RR_{\lambda''}$,
\begin{eqnarray}
I_1=\sum_\ell f(\RR_\ell)f(\RR_\ell-\Delta\RR_{\lambda''})
e^{-i\kk_\mu\cdot\RR_\ell},
\label{eqI_1I_1}
\end{eqnarray}
where the even function $f(\XX_\ell)$ is defined as
\begin{eqnarray}
 f(\XX_\ell)=\cos(\kk_{\lambda'}\cdot\XX_\ell)\,e^{{-|\XX_l|}/{L_{\lambda'}}}.
\label{eqI_1I_2}
\end{eqnarray}
Since the localization lengths of SL modes are the same, $e.g.$, $L_{\lambda'}\cong L_{\lambda''}$, hereafter we denote this as $L$.
As $f(\XX_\ell)$ is concerned with SL modes, the relevant sum should be made in the region  $\mid\XX_\ell\mid\leq L$, so we can approximate the summation by
\begin{equation}
\begin{split}
I_1&\cong\frac{1}{\Omega}\int_{\mid\XX_\ell\mid<L}d\RR_\ell f(\RR_\ell)f(\RR_\ell-\Delta\RR_{\lambda''})
e^{-i\kk_\mu\cdot\RR_\ell}
\\
&\cong\frac{1}{\Omega}\int_{\mid\XX_\ell\mid<L}d\RR_\ell\, e^{\frac{{-|\RR_\ell|}-{|\RR_\ell-\Delta\RR_{\lambda''}|}}{L}}e^{-i\kk_\mu\cdot\RR_\ell}\\
&\left[\frac{1}{2}\cos\left(2\kk_{\lambda'}\cdot\RR_\ell-\kk_{\lambda'}\cdot\Delta\RR_{\lambda''}\right)+\frac{1}{2}\cos\left(\kk_{\lambda'}\cdot\Delta\RR_{\lambda''}\right)\right]
\\
&\cong|\Delta \RR_{\lambda''}|\pi L^2\frac{1}{2\Omega}e^{-|\Delta\RR_{\lambda''}|/L}.
\end{split}
\label{eqI_1I_1}
\end{equation}
where we have used the approximation $\cos(\kk_{\lambda'}\cdot\Delta\RR_{\lambda''})\approx\cos( k_{\lambda'} nL)\approx 1$ from the Ioffe-Regel condition $L\approx 2\pi/k_{\lambda'}$ for SL modes and $e^{-i\kk_\mu\cdot\RR_\ell}\approx 1$ due to $\mid\kk_\mu\mid\ll 2\pi/L$ for the wave number of EX acoustic modes. The term containing $\cos\left(2\kk_{\lambda'}\cdot\RR_\ell-\kk_{\lambda'}\cdot\Delta\RR_{\lambda''}\right)$ becomes negligible since it yields rapidly oscillating function in the integrand.

%\begin{eqnarray}I_1&\simeq&\red{8}\sum_{\ell=0}^{\infty}e^{-\RR_l\cdot\hat{\bm{L}}_{\lambda'}/L_{\lambda'}}e^{-(\RR_l+\Delta\RR_{\lambda', \lambda''})\cdot\hat{\bm{L}}_{\lambda''}/L_{\lambda''}}e^{-i\kk_\mu\cdot\RR_l}.\nonumber\\&\simeq& \frac{\red{8}}{\Omega}\prod_{\alpha=x,y,z}\frac{1}{2/L_{\lambda}^\alpha+ik_\mu^\alpha}e^{-\Delta\RR_{\lambda', \lambda''}\cdot\hat{\bm{L}}_{\lambda''}/L_{\lambda''}}.\label{eq32I56}\end{eqnarray}
This gives the squared hopping integral of the form
\begin{eqnarray}
I_1^2\simeq \left( \frac{\pi\Delta R_{\lambda''} L^2}{2\Omega}\right)^2 e^{-2\Delta R_{\lambda''}/L},
\label{eq32I66}
\end{eqnarray}
where $\Delta R_{\lambda''}$ is the hopping distance.
%We emphasize that the hopping distance should be the nearest neighbor position $\Delta R_{\lambda''}\approx a_0/2$ and the localization length is comparable with the wave length, so $L_{\lambda}\approx a_0$.
%As a result, the exponent in Eq.\,(\ref{eq32I66}) becomes $\approx e^{-1}$.
%Since the localization lengths of SL modes are the same, $e.g.$, $L_{\lambda'}\cong L_{\lambda''}$, hereafter we denote this as $L$.
%%%%%%%%%%%%%%
%%%%%%%%%%%%%%%%%%%%%%

In the temperature regime {\it T}$\simeq$\,a few 10\,K, $i.e.$, $k_BT> \hbar\omega_{\lambda'}, \hbar\omega_{\lambda''}> \hbar\omega_{\kk\mu}$, the inverse of the relaxation time takes the following form under the above conditions and by employing the linear dispersion relation for EX phonon mode $\omega_{\kk\mu}=v_\mu k_\mu$,
\begin{small}
\begin{equation}
\begin{split}
\frac{1}{\tau_{\lambda'}^{\rm same}}
&\cong
\frac{2\pi k_BT(D_r+D_\phi)^2CI_1^2}{\hbar^3VL^6v_\parallel^2}
\\
&\times\sum_{\kk_\parallel,{\lambda''}}\left[ \frac{\delta(\omega_{\lambda''}-\omega_{\lambda'}-\omega_{\kk_\parallel})}{\omega_{\lambda''}^2}
+\frac{\delta(\lambda'\rightleftarrows\lambda'')}{\omega_{\lambda''}^2}\right]
\\
&+\left[2\times\left(D\longrightarrow S, \parallel\longrightarrow \perp \right) \,\,\rm{in\,the\,above} \right].
%&+2\frac{2\pi k_BT(S_r+S_\phi)^2CI_1^2}{\hbar^3VL^6v_\perp^2}\sum_{\kk_\perp,{\lambda''}}\left[\frac{\delta(\omega_{\lambda''}-\omega_{\lambda'}-\omega_{\kk_\perp})}{\omega_{\lambda''}^2}+\frac{\delta(\lambda'\rightleftarrows\lambda'')}{\omega_{\lambda''}^2}\right].
%\\
%&+\frac{2\pi(D_r+D_\phi)^2k_\parallel^2CI_1^2}{\hbar^2VL^6}\sum_{\kk_\parallel,{\lambda''}}\frac{1}{\omega_{k_\parallel}\omega_{\lambda'}\omega_{\lambda''}}\delta(\omega_{\lambda'}-\omega_{\lambda''}-\omega_{k_\parallel})
%\\
%&+2\frac{2\pi(S_r+S_\phi)^2k_\perp^2CI_1^2}{\hbar^2VL^6}\sum_{\kk_\perp,{\lambda''}}\frac{1}{\omega_{k_\perp}\omega_{\lambda'}\omega_{\lambda''}}\delta(\omega_{\lambda'}-\omega_{\lambda''}-\omega_{k_\perp})
\label{eq60}
\end{split}
\end{equation}
\end{small}
Here the coefficient $C$ is defined in Eq.\,(\ref{eq51}). We have omitted the temperature independent term providing  only small contributions.
%The last two terms in Eq.\,\ref{eq60}compared with the first two terms . So we neglect these terms hereafter.
%%%%%%%%%%%%
%%%%%%%%%%%%%%%%%%
%%%%%%%%%%%%%%%%%%%%%%%
\subsection{Thermal conductivity due to the hopping  of SL modes}
In the previous subsection, we have formulated the relaxation rate of SL modes due to the anharmonic interaction between SL modes and EX modes.
This is a quantum process realizing the decay of SL$'$ mode to SL$''$ mode assisted by EX mode: SL$'$+EX$\rightarrow$SL$''$.
Without anharmonic interaction, SL modes cannot diffuse/contribute to thermal transport.
This means that the plateau region should continue over at higher temperatures after exhibiting the onset of the plateau, $i.e.$, the contribution from EX modes to lattice thermal conductivity is saturated at higher temperatures.
This is because the onset of the plateau arises from the weak localization of acoustic modes as explained in Sec.II.
Thus, the {\it T}-linear rise of $\kappa_{\rm L}(T)$ cannot recover without anharmonic interaction between SL modes and EX modes.

In addition, we emphasize that disorder, induced by off-centeredness as shown in Supplemental Material, is essential to
generate the hopping of SL modes.
This occurs only in the case that SL$'$ mode belonging to the eigenfrequency $\omega_{\rm SL'}$ can hop to a  site of SL$''$ mode with a different eigenfrequency $\omega_{\rm SL''}$ via absorption or emission of EX mode with finite frequency $\pm(\omega_{\rm SL'}-\omega_{\rm SL''})$.
This finite frequency is created by level repulsion between eigenfrequencies due to disorder, $i.e.$, localized modes never belong to the same eigenfrequency according to the level repulsion.

Let us provide the formula of $\kappa_{\rm L}(T)$ due to the diffusion process where
SL modes serve as primary heat carriers.
In this process, the characteristic length-scale should be the hopping distance $\Delta R_{\lambda''}$ from the site of SL$'$ mode to that of SL$''$ mode,
and the characteristic time-scale is the relaxation time $\tau_{\lambda'}$ of the SL$'$ mode.
This leads to the following formula of the lattice thermal conductivity due to the hopping process, which was first proposed for fracton excitations by Alexander $et.~al.$\,\cite{Alexander1987},

%{When SL modes are fully excited, the contribution from EX modes to thermal conductivity is negligible due to the scattering from Umklapp process and hopping process. In this case, SL modes will serve as primary heat carriers. With the single mode relaxation time estimated in the previous section, the thermal conductivity related with the diffusion of SL modes due to hopping process is expressed by\,\cite{Alexander1987}}
%The formula of thermal conductivity due to the hopping of SL modes is expressed by\,\cite{Alexander1987}
\begin{equation}
\begin{split}
\kappa_{\rm hop}(T)=\frac{1}{3V}\sum_{\lambda'}C_{\lambda'}(T)\frac{\Delta R_{\lambda''}^2}{\tau_{\lambda'}},
\end{split}
\label{eq61234}
\end{equation}
where $\Delta R_{\lambda''}^2/\tau_{\lambda'}$ is the thermal diffusivity of SL mode $\lambda'$, $C_{\lambda'}(T)$ is the specific heat associated with the SL mode $\lambda'$.
In the high temperature regime above the plateau region {\it T}$\gtrsim$a few 10\,K, the specific heat follows the Dulong-Petit relation of the form $C_{\lambda'}(T)=k_{\mathrm B}$ per one polarization of SL mode $\lambda'$. Note that $1/\tau_{\lambda'}=1/\tau_{\lambda'}^{\rm same}+1/\tau_{\lambda'}^{\rm dif}$, we first calculate the hopping process between the same polarization by,
%%%%%%%%%%%%%%%%%%%%%%%%
\begin{equation}
\begin{split}
\kappa_{\rm hop}^{\rm same}(T)=\frac{k_B}{3V}\sum_{\lambda'}\frac{\Delta R_{\lambda''}^2}{\tau_{\lambda'}^{\rm same}}.
\end{split}
\label{eq6123}
\end{equation}
%%%%%%%%%%%%%%%%%%%
%%%%%%%%%%%%%%%%%%%%%%
The substitution of Eq.\,(\ref{eq60}) into Eq.\,(\ref{eq6123}) together with
%\begin{small}\begin{equation}\begin{split}
%\kappa_{\rm hop}^{\rm same}(T)&\cong\frac{k_B}{3V^2}\frac{2\pi k_BT(D_r+D_\phi)^2C}{\hbar^3v_\parallel^2L^6}\sum_{\kk_\parallel,\lambda',\lambda''}I_1^2\frac{\Delta R_{\lambda''}^2}{\omega_{\lambda''}^2}
%\\&\times\left[\delta(\omega_{\lambda''}-\omega_{\lambda'}-\omega_{\kk,\parallel})+\delta(\lambda'\rightleftarrows\lambda'')\right]
%\\&+2\frac{k_B}{3V^2}\frac{2\pi k_BT(S_r+S_\phi)^2C}{\hbar^3v_\perp^2L^6}\sum_{\kk_\perp,\lambda'',\lambda'}I_1^2\frac{\Delta R_{\lambda''}^2}{\omega_{\lambda''}^2}
%\\&\times\left[\delta(\omega_{\lambda''}-\omega_{\lambda'}-\omega_{\kk,\perp})+\delta(\lambda'\rightleftarrows\lambda'')\right].
%&+\frac{k_B}{3V^2}\frac{2\pi (D_r+D_\phi)^2k_\parallel^2C}{\hbar^2L^6}\sum_{\kk_\parallel,\lambda'',\lambda'}I_1^2\frac{\Delta R_{\lambda''}^2}{\omega_{k_\parallel}\omega_{\lambda'}\omega_{\lambda''}}
%\\&\times\delta(\omega_{\lambda'}-\omega_{\lambda''}-\omega_{\kk,\parallel})\\
%&+2\frac{k_B}{3V^2}\frac{2\pi (S_r+S_\phi)^2k_\perp^2C}{\hbar^2L^6}\sum_{\kk_\perp,\lambda'',\lambda'}I_1^2\frac{\Delta R_{\lambda''}^2}{\omega_{k_\perp}\omega_{\lambda'}\omega_{\lambda''}}
%\\&\times\delta(\omega_{\lambda'}-\omega_{\lambda''}-\omega_{\kk,\perp})
%\end{split}\end{equation}\end{small}
%%%%%%%%%%%%
%%%%%%%%%%%%%%%
Eq.\,(\ref{eq32I66}) yields
\begin{small}
%\begin{eqnarray}
\begin{equation}
\begin{split}
\kappa_{\rm hop}^{\rm same}(T)
&\cong\frac{k_B}{3V^2}\frac{\pi^3 k_BT(D_r+D_\phi)^2C}{2\hbar^3v_\parallel^2L^2\Omega^2}\sum_{\kk_\parallel,\lambda',\lambda''}\frac{\Delta R_{\lambda''}^4}{\omega_{\lambda''}^2}
\\
&\times
e^{-2\Delta R_{\lambda''}/L}
\left[\delta(\omega_{\lambda''}-\omega_{\lambda'}-\omega_{\kk,\parallel})+
\delta(\lambda'\rightleftarrows\lambda'')\right]
\\
&+\left[2\times\left(D\longrightarrow S, \parallel\longrightarrow \perp \right) \,\,\rm{in\,the\,above} \right]
%\\
%&+2\frac{k_B}{3V^2}\frac{\pi^3 k_BT(S_r+S_\phi)^2C}{2\hbar^3v_\perp^2L^2\Omega^2}\sum_{\kk_\perp,\lambda'',\lambda'}\frac{\Delta R_{\lambda''}^4}{\omega_{\lambda''}^2}
%\\
%&\times
%e^{-2\Delta R_{\lambda''}/L}\left[\delta(\omega_{\lambda''}-\omega_{\lambda'}-\omega_{\kk,\perp})+\delta(\lambda'\rightleftarrows\lambda'')\right]
%\\&+\frac{k_B}{3V^2}\frac{\pi^3 (D_r+D_\phi)^2k_\parallel^2C}{2\hbar^2L^2\Omega^2}\sum_{\kk_\parallel,\lambda'',\lambda'}\frac{\Delta R_{\lambda''}^4}{\omega_{k_\parallel}\omega_{\lambda'}\omega_{\lambda''}}
%\\&\timese^{-2\Delta R_{\lambda''}/L}\delta(\omega_{\lambda'}-\omega_{\lambda''}-\omega_{\kk,\parallel})
%\\&+2\frac{k_B}{3V^2}\frac{\pi^3 (S_r+S_\phi)^2k_\perp^2C}{2\hbar^2L^2\Omega^2}\sum_{\kk_\perp,\lambda'',\lambda'}\frac{\Delta R_{\lambda''}^4}{\omega_{k_\perp}\omega_{\lambda'}\omega_{\lambda''}}
%\\&\timese^{-2\Delta R_{\lambda''}/L}\delta(\omega_{\lambda'}-\omega_{\lambda''}-\omega_{\kk,\perp})
\end{split}
%\end{eqnarray}
\end{equation}
\end{small}
%%%%%%%%%%%%%%%%%%%%%%%%%%%%%%%%%%%%%%%%%

Transforming the sum $\sum_{\kk_\mu}$ for EX phonon modes  to  the integral $V/(2\pi)^3\int d\kk_\mu=V/(2\pi^2v_\mu^3)\int \omega_{k_\mu}^2d\omega_{k_\mu}$,
%by employing the linear dispersion relation  $\omega_{k\mu}=v_\mu k_\mu$,
we have
\begin{equation}
\begin{split}
&\kappa_{\rm hop}^{\rm same}(T)
=\frac{\pi k_B^2TC}{12\hbar^3V\Omega^2L^2}\left[\frac{(D_r+D_\phi)^2}{v^5_\parallel}+2\frac{(S_r+S_\phi)^2}{v^5_\perp}\right]
\\
&\times
\sum_{\lambda'',\lambda'}\Delta R_{\lambda''}^4e^{-2\Delta R_{\lambda''}/L}
\frac{(\omega_{\lambda''}-\omega_{\lambda'})^2}{\omega_{\lambda''}^2}
%\\&+\frac{\pi k_BC}{12\pi\hbar^2V\Omega^2L^2}\left[\frac{(D_r+D_\phi)^2}{v^5_\parallel}+2\frac{(S_r+S_\phi)^2}{v^5_\perp}\right]
%\\&\times\sum_{\lambda'',\lambda' \atop \omega_{\lambda'}>\omega_{\lambda''}}\Delta R_{\lambda''}^4e^{-2\Delta R_{\lambda''}/L}\frac{(\omega_{\lambda'}-\omega_{\lambda''})^3}{\omega_{\lambda'}\omega_{\lambda''}}.
\label{eq:dif1}
\end{split}
\end{equation}
The sum on $\lambda'$ and $\lambda''$ above should include the density of states of SL modes $D_{\rm SL}(\omega_{\lambda'})$ and $D_{\rm SL}(\omega_{\lambda''}(\Delta R_{\lambda''}))$ for the same polarization process.
The volume $\Omega$ should contain two independent SL modes corresponding to two independent in-plane mode, say, stretching or libration, in the band width of $\Delta\omega_{\rm sl}$, which leads to
\begin{equation}
\begin{split}
D_{\rm SL}(\omega_{\lambda'})\Omega\Delta\omega_{\rm sl}=2.
\label{eq:dif0}
\end{split}
\end{equation}
and
\begin{equation}
\begin{split}
D_{\rm SL}(\omega_{\lambda''}(\Delta R_{\lambda''}))\Omega\Delta\omega_{\rm sl}=1.
\end{split}
\label{eq:dif1}
\end{equation}
%%%%%%%%%%%%%%%%%%%%%%%%%%%%%%%%%
where the volume $\Omega$ contains at least one possible SL mode $\lambda''$ with the same/different polarization as/from mode $\lambda'$.
Since the term $\Delta R_{\lambda''}^4e^{-2\Delta R_{\lambda''}/L}$ in Eq.\,(\ref{eq:dif1}) achieves its maximum at $\Delta R_{\lambda''}=2L$ and it decays fast with the further increasing of $\Delta R_{\lambda''}$, the sum of $\lambda''$ could be estimated within the sphere region $\Delta R_{\lambda''}\leq\Delta R$.
\begin{equation}
\begin{split}
&\sum_{\lambda'',\lambda'}\Delta R_{\lambda''}^4e^{-2\Delta R_{\lambda''}/L}
\frac{(\omega_{\lambda''}-\omega_{\lambda'})^2}{\omega_{\lambda''}^2}
%\\
%&=\frac{V}{\Omega\Delta\omega_{sl}}\int_{\omega_{sl}}^{\omega_{sl}+\Delta\omega_{sl}}d\omega_{\lambda'}
%\sum_{\lambda''}\Delta R_{\lambda''}^4e^{-2\Delta R_{\lambda''}/L}
%\frac{(\omega_{\lambda''}-\omega_{\lambda'})^2}{\omega_{\lambda''}^2}
%\\
%&\cong
%\frac{4\pi}{3}(2L)^3\frac{V}{\Omega^2\Delta\omega_{sl}^2}(2L)^4e^{-4}
%\\
%&\times\int_{\omega_{sl}}^{\omega_{sl}+\Delta\omega_{sl}}d\omega_{\lambda'}
%\int_{\omega_{sl}}^{\omega_{sl}+\Delta\omega_{sl}}d\omega_{\lambda''}\frac{(\omega_{\lambda''}-\omega_{\lambda'})^2}{\omega_{\lambda''}^2}
\\
&\cong\frac{\frac{4\pi}{3}\Delta R^3 2V}{\Omega^2}\Delta R^4e^{-2\Delta R/L}\times(10^{-2})
\label{eq:dif00}
\end{split}
\end{equation}
%and
%\begin{equation}
%\begin{split}
%\red{??\sum_{\lambda'',\lambda' \atop \omega_{\lambda'}>\omega_{\lambda''}}\Delta R_{\lambda''}^4e^{-2\Delta R_{\lambda''}/L}
%\frac{(\omega_{\lambda'}-\omega_{\lambda''})^3}{\omega_{\lambda'}\omega_{\lambda''}}
%}
%\end{split}
%\end{equation}
Here the sum on SL modes are done by $\sum_{\lambda''}=4\pi\Delta R^3/3\int_{\omega_{\rm sl}}^{\omega_{\rm sl}+\Delta\omega_{\rm sl}}
D(\omega_{\lambda''}(\Delta R_{\lambda''}))d\omega_{\lambda''}$ and $\sum_{\lambda'}=V\int_{\omega_{\rm sl}}^{\omega_{\rm sl}+\Delta\omega_{\rm sl}}
D(\omega_{\lambda'})d\omega_{\lambda'}$, where the factor $4\pi\Delta R^3/3\Omega$ from Eq.\,(\ref{eq:dif1}) means the total number of hopping sites from $\lambda'$ to $\lambda''$ for the same polarization process, and $2V/\Omega$ from Eq.\,(\ref{eq:dif0}) is the total number of $\lambda'$ contributing the thermal conductivity $\kappa_{\rm hop}$.
The numerical factor $10^{-2}$ arises from the magnitude estimation of integral $\int_{\omega_{\rm sl}}^{\omega_{\rm sl}+\Delta\omega_{\rm sl}}d\omega_{\lambda'}
\int_{\omega_{\rm sl}}^{\omega_{\rm sl}+\Delta\omega_{\rm sl}}d\omega_{\lambda''}\frac{(\omega_{\lambda''}-\omega_{\lambda'})^2}{\Delta\omega_{\rm sl}^2\omega_{\lambda''}^2}
 $.

The formula of the thermal conductivity due to the hopping mechanism is given by
\begin{equation}
\begin{split}
\kappa_{\rm hop}^{\rm same}(T)=&\frac{\pi^2k_B^2 T\Delta R^7}{144\rho m'^2\Omega^2 L^2 }e^{-2\Delta R/L}\times(10^{-2})
\\
&\left[\frac{(D_r+D_\phi)^2}{v^5_\parallel}+2\frac{(S_r+S_\phi)^2}{v^5_\perp}\right]
\label{eqdif2}
\end{split}
\end{equation}
%%%%%%%%%%%%%%%%
%%%%%%%%%%%%%%%%

The same procedure for the hopping process due to anharmonic interaction between different polarizations
leads to
%as Eq.\,(\ref{eqdif2}) is obtained, $i.e.$, $\kappa_{\rm hop}^{xx}(T)=\kappa_{\rm hop}^{yy}(T)$.
%%%%%%%%%%%%%%%%%%%%%%%%%%%%
%\begin{equation}\kappa_{\rm hop}^{yy}(T)=\frac{k_B^2 Ta^2}{3\pi^3\Omega\rho m'^2 }e^{-2a/L}\left[\frac{D^2}{v^5_\parallel}+2\frac{S^2}{v^5_\perp}\right]\label{eqhop2}\end{equation}
%%%%%%%%%%%%

%%%%%%%%%%%%%%%%%%%%%%%%
\begin{equation}
\begin{split}
\kappa_{\rm hop}^{\rm dif}(T)=&\frac{4^2k_B^2 T\Delta R^7}{144\rho m'^2\Omega^2 L^2 }e^{-2\Delta R/L}\times(10^{-2})
\\
&\left[\frac{(D_r-D_\phi)^2}{v^5_\parallel}+2\frac{(S_r-S_\phi)^2}{v^5_\perp}\right]
\label{eqhop3}
\end{split}
\end{equation}
The total thermal conductivity due to the hopping mechanism is given by the sum of these components as
\begin{equation}
\kappa_{\rm hop}(T)=\kappa_{\rm hop}^{\rm same}(T)+\kappa_{\rm hop}^{\rm dif}(T),
\label{eqdif_total_DS}
\end{equation}
%where we have employed the relation $L^3/\Omega\cong a_0^3/(a_0/2)^3\cong 2^3$.
\\
%%%%%%%%%%
%The integral could be solve analytically,
%\begin{equation}
%\begin{split}
%&\int_0^{k_c} dkk^2\left[\frac{1}{(b+k)^2}+\frac{1}{(b-k)^2}\right]
%\\
%&=\left[-2b\ln(|k+b|)+2b\ln(|k-b|)-\frac{b^2}{k+b}-\frac{b^2}{k-b}+2k\right]_0^{k_c}
%\\
%&=2b\ln\frac{b-k_c}{b+k_c}-2k_c\frac{b^2}{k_c^2-b^2}+2k_c
%\end{split}
%\end{equation}
%and,
%\begin{equation}
%\begin{split}
%\frac{1}{\tau_{\lambda'}}
%&=\frac{\pi k_BT D^2L^2}{\hbar^3 v_\parallel^4}C\,\overline{N}_{\rm SL}\frac{V}{(2\pi)^3}2\pi\left[2b_\parallel\ln\frac{b_\parallel-k_c}{b_\parallel+k_c}-2k_c\frac{b_\parallel^2}{k_c^2-b_\parallel^2}+2k_c\right]
%\\
%&+2\frac{\pi k_BT R^2L^2}{\hbar^3 v_\perp^4}C\,\overline{N}_{\rm SL}\frac{V}{(2\pi)^3}2\pi\left[2b_\perp\ln\frac{b_\perp-k_c}{b_\perp+k_c}-2k_c\frac{b_\perp^2}{(\pi/2a)^2-b_\perp^2}+2k_c\right]
%\end{split}
%\end{equation}
%where $b_\mu=\omega_{\lambda'}/v_\mu$.

\subsection{Evaluation of anharmonic coupling $D$ and $S$}
Here we estimate the anharmonic coupling constants $D_{r(\phi)}$ and $S_{r(\phi)}$
by illustrating type-I BGS.
%from \red{both the data of pressure dependence of Raman\,\cite{Kume2015}
%and infrared\,\cite{Mori2011}spectroscopies and the first-principles phonon calculation} on relevant modes for off-center type-I BGS.
The coupling constants $D_r(S_r)$ and $D_\phi(S_\phi)$ are associated with the stretching and libration motion of guest-cage vibrations identified by the force constant $\xi_r$ and $\xi_\phi$   in Eq.\,(\ref{eq88vec}) by the relation $\xi_{r(\phi)}=m'\omega_{r(\phi)}^2$, where $m'$ is the reduced mass defined by $1/m'=1/m+1/M$.
In our coarse-grained Hamiltonian introduced in Sec.\,III, the guest ion Ba(2) in tetrakaidecahedron cage has the mass $m$ and the molecular unit
composed of 1 tetrakaidecahedron and 1/3 dodecahedron does the total mass $M$ excluding the off-center guest ion.

We first evaluate the coupling constants  $D_{r(\phi)}$ from the Raman spectroscopy data of pressure dependence \,\cite{Kume2015}. The $D_r$ can be related to the pressure $P$ by
\begin{equation}
\begin{split}
D_r=\frac{\partial\xi_r}{\partial e_{\alpha\alpha}}
=3B\frac{\partial\xi_r}{\partial \omega^r}\frac{\partial \omega^r}{\partial P}
=3B(2m'\omega_0^r)\frac{\partial \omega^r}{\partial P}.
%%%%%%%%%%%%%
%=\frac{\partial\xi_r}{\partial T}\frac{\partial T}{\partial e_{\alpha\alpha}}=\frac{1}{\beta_{\alpha\alpha}}\frac{\partial\xi_r}{\partial T}.
\end{split}
\label{eqDD1}
\end{equation}
Here $B=\frac{\Delta P}{(\Delta V/V)}$ is the linear thermal expansion coefficient, where
the dilation is given by
%$\delta=
$\Delta V/V=\sum_\alpha e_{\alpha\alpha}$ for cubic structure.
%Equation (\ref{eqDD1}) can be expressed by
%\begin{equation}
%D_r=\frac{1}{\beta_{\alpha\alpha}}\frac{\partial\xi_r}{\partial \omega^r}\frac{\partial \omega^r}{\partial T}
%=\frac{1}{\beta_{\alpha\alpha}}(2m'\omega_0^r)\frac{\partial \omega^r}{\partial T}.
%\label{eqDDD1}\end{equation}
The coupling constant $D_\phi$ can be defined in a similar manner to Eq.\,(\ref{eqDD1}) as
\begin{equation}
\begin{split}
D_\phi
&=\frac{\partial\xi_\phi}{\partial e_{\alpha\alpha}}
%=\frac{1}{\beta_{\alpha\beta}}\frac{\partial\xi_\phi}{\partial T}
%=\frac{1}{\beta_{\alpha\beta}}\frac{\partial\xi_\phi}{\partial \omega}\frac{\partial \omega}{\partial T}
=3B(2m'\omega_0^\phi)\frac{\partial \omega^\phi}{\partial P}.
\end{split}
\label{eqSS1}
\end{equation}
%These relations yield the anharmonic coupling constants $D_r$ and $D_\phi$ below.
In the pressure range from 0.8 GPa to 5.8 GPa, E$_{\rm g}$ mode spans from ~20 cm$^{-1}$ to ~27\,cm$^{-1}$.
While, for T$_{\rm 2g}$ mode, it ranges from ~17 cm$^{-1}$ to ~27 cm$^{-1}$.
The observed spectra of these two modes are overlapped/mixed.
Taking account of these aspects, we have
%We note that the same procedure to determine the force constants from the data of pressure dependence has been adopted for structural glasses.\cite{Nakayama1999}
%By extrapolating these data, the mode frequency of the A$_{1\mathrm{g}}$ mode of BGS is estimated as $\nu_0$ = 30\,cm$^{-1}$.
$\partial \omega^r/\partial P=2\pi\times4.2\times 10^{10}\,\rm{[sec^{-1}GPa^{-1}]}$ and $\partial \omega^\phi/\partial P=2\pi\times6.0\times 10^{10}\,\rm{[sec^{-1}GPa^{-1}]}$.
We then obtain the coupling constants  $D_r=m'\pi^2\times 3.0\times 10^{25}\,\rm{[kg\cdot sec^{-2}]}$ and $D_\phi
=m'\pi^2\times 3.0\times 10^{25}\,\rm{[kg\cdot sec^{-2}]}$ using the observed bulk modulus $B=41.3 \rm{GPa}$\,\cite{Ishii2012}.
Within our knowledge, the experiment data for estimating the coupling coefficients $S_{r(\phi)}$ are not available, so we assume as $S_r\approx D_r$ and $S_\phi\approx D_\phi$ at the present stage.
The above coupling constants yield
\begin{equation}
\kappa_{\rm hop}=3.3\times10^{-3}T (\mathrm{Wm^{-1}K^{-1}}),
\label{eq:Raman_kappa1}
\end{equation}
where we have employed the values of parameters in Eq.\,(\ref{eqdif_total_DS})
as the localization length $L=2a_0$, the hopping distance $\Delta R=3.5L$, the volume of molecular unit $\Omega=(a_0)^3/6$, the lattice spacing $a_0=11.68$\AA, the mass density $\rho=6.01\times 10^3$\,kg/m$^3$, in addition to the velocities of acoustic phonons $v_\parallel=3369$\,m/s and $v_\perp=1936$\,m/s\,\cite{Avila:2006a}.
The value of $\kappa_{\rm hop}$ in Eq.\,(\ref{eq:Raman_kappa1}) is smaller than the observed one of $\kappa_{\rm hop}=9.2\times10^{-3}T (\mathrm{Wm^{-1}K^{-1}})$ for type-I BGS.
This mainly arises from, as will be demonstrated below by means of FPC, the underestimated shear coupling constants $S_{r(\phi)}$ obtained by assuming the relations $S_{r(\phi)}\approx D_{r(\phi)}$.
%which is slightly smaller than the value of $\kappa_{\rm hop}$ in Eq.\,(\ref{eq:fin_kappa1}).

Due to the lack of experiment data for the shear coupling coefficients $S_{r(\phi)}$,
we have performed FPC for type-I BGS to obtain the coupling constants from the shift of eigenfrequencies at $\Gamma$-point of low-lying optical mode by imposing strain to the cage structure.
The normal strain is isotropic and defined as $e_{\alpha\alpha}=(a_0-a)/a_0$ where $a_0$ and $a$ are the lattice constant for the unstrained and strained unit cell\,\cite{Chen2014}, respectively.
The shear strain is also isotropic and defined as $e_{\alpha\beta}=(1-\sqrt{1-(2\cos\theta-1)\cos\theta})/(2\cos\theta-1)$ where $\theta$ is the acute angle between edges after deformation.

We have performed the FPC by the VASP code\,\cite{Kresse1999} with the Perdew-Burke-Ernzerhof functional and the PAW method\,\cite{Perdew1996}, plane wave cut-off energy 250 eV and the force convergence less than $10^{-7} \rm{eV/\AA}$.
The phonon frequencies are calculated by PHONOPY code\,\cite{Togo2015} with the $4\times4\times4$ Monkhorst-Pack $k$ grids and for a unit cell containing 54 atoms.
The coupling constants obtained from normal strain are
$D_r=m'\pi^2\times 2.1\times 10^{25}\,\rm{[kg\cdot sec^{-2}]}$, $D_{\phi}=m'\pi^2\times 1.5\times 10^{25}\,\rm{[kg\cdot sec^{-2}]}$, and from sheared unit cell are $S_r=m'\pi^2\times 4.2\times 10^{25}\,\rm{[kg\cdot sec^{-2}]}$, $S_{\phi}=m'\pi^2\times 2.9\times 10^{25}\,\rm{[kg\cdot sec^{-2}]}$, respectively.
The $D_{r(\phi)}$ are smaller than those estimated from the Raman spectroscopy data of pressure dependence,
though  $S_{r(\phi)}$ are larger than the values obtained from the assumption
$S_{r(\phi)}\approx D_{r(\phi)}$.
%Moreover, these results indicate that low-lying optical modes are sensitive to the shear distortion.
%Detailed calculations will be published elsewhere.\cite{FPC2017}
The above coupling constants yield the thermal conductivity due to the hopping of SL modes
of
\begin{equation}
\kappa_{\rm hop}=4.8\times10^{-3}T (\mathrm{Wm^{-1}K^{-1}}).
\label{eq:fin_kappa1}
\end{equation}
We remark here that our FPC provides the results for the on-center positioned Ba(2)
because the optimization for off-center structure is quite time-consuming
and may require to take into account the dipole-dipole interaction due to off-centeredness and temperature effect.
The on-center structure gives rise to the underestimated coupling constants $S$ since on-center guest ions should more weakly response to shear distortion than the case of off-center.
Then, the actual $S_{r(\phi)}$ should be larger than the above estimation.
%Detailed calculations on this aspect will be published elsewhere.
Under these situations, the calculated value in Eq.\,(\ref{eq:fin_kappa1}) provides reasonable agreement, to claim the relevance of the hopping process of SL modes, with the observed $\kappa_{\rm hop}= \gamma T$ with $\gamma=9.2\times 10^{-3}\,\mathrm{Wm^{-1}K^{-2}}$ for type-I  BGS\,\cite{Avila:2006a, Avila:2008aa}, and $\gamma=9.0\times 10^{-3}\,\mathrm{Wm^{-1}K^{-2}}$ for type-I EGG\,\cite{Sales:2001a}.
For type-I SGG, several different values around $\gamma\sim 8.0\times10^{-3}\,\mathrm{Wm^{-1}K^{-2}}$ have been reported\,\cite{Nolas:1998aa, Cohn:1999a, Christensen:2016a, Suekuni:2007a}, indicating that the experimental data  of SGG depend on sample qualities according to synthesis methods.
In that respect, it has been reported\,\cite{Christensen:2016a} that a flux-grown sample shows a
glasslike plateau, while a zone-melted sample has a crystalline peak.

\section{Summary and Conclusions}
Off-center type-I clathrates  show almost identical lattice thermal conductivities $\kappa_{\rm L}$ to those of structural glasses %when off-center guest ions take disordered orientation, though their scaffold caged-network takes crystalline cubic-structure
\,\cite{Nolas:1998aa, Cohn:1999a, Sales:2001a, Paschen:2001a, Suekuni:2007a, Avila:2008aa, Suekuni:2008a}.
In addition, off-center type-I clathrates show the excess density of states at THz frequencies manifesting the boson peak identical to those of network-forming  glasses\,\cite{Avila:2006a, Avila:2008aa, Suekuni:2008a}.
These indicate that the symmetry broken guest ions  in cages take charge of the emergence of glasslike  $\kappa_{\rm L}(T)$.
In structural glasses, many key aspects of a detailed quantitative description are still missing.
%the underlying mechanisms on lattice thermal conductivities
% including the plateau at around several K and the subsequent rise proportional to T have not yet been fully understood.
This is due to the difficulty to identify relevant entities or elements at atomic scale caused by their complex microscopic structures.
%except a few cases, $e.g.$, M$_2$O$\cdot n$B$_2$O$_3$ glasses (M=Li$^+$ or Ag$^+$, and $n$ is the molecular number), where a role of weakly coupled rattling ions for the boson peak is investigated in terms of Raman scattering measurements\,\cite{D'Angelo2011}.

%Liu $et. al.$~\cite{Liu2016} have recently investigated,  in terms of large-scale numerical simulations,  peculiar THz frequency dynamics of off-center type-I clathrate compounds. They have found that there appear three kinds of modes at THz frequency region classified into extended (EX), weakly (WL) and strongly localized (SL) modes in the THz frequency region and below. These modes are associated with the hybridization of acoustic modes relevant to network cages with local modes of off-center guest ions\,\cite{Liu2016}.
%In this paper, we have fully incorporated the above findings in order to identify the underlying mechanism of glasslike  $\kappa_{\rm L}$ of off-center type-I clathrates.

In Sec.\,II, we have pointed out that the PR shown in Fig. \ref{fig33}  provides the evidence that EX acoustic phonons carrying heat convert to WL modes modes at $\sim$\,1.3\,meV in off-center BGS.
This energy corresponds to the temperature 3.9\,K\,$\approx$1.3\,meV/3.84k$_{\rm{B}}$ from the Wien's displacement law, so that this conversion should be associated with the onset of the plateau thermal conductivities observed at several K in off-center type-I clathrates~\cite{Nolas:1998aa, Cohn:1999a, Sales:2001a, Paschen:2001a, Avila:2006a, Suekuni:2007a, Avila:2008aa, Suekuni:2008a}.

With increasing temperature further, thermal conductivities above a few 10\,K show a linear rise on temperature.
This type of anomalous thermal conductivities with the plateau and the subsequent {\it T}-linear rise have been clearly observed   for off-center type-I clathrates~\cite{Nolas:1998aa, Cohn:1999a, Sales:2001a, Paschen:2001a, Avila:2006a, Suekuni:2007a, Avila:2008aa, Suekuni:2008a}.
This is the prominent hallmark of glasslike thermal conductivity
since crystals with translational invariance never show these features.
Rather, lattice thermal conductivities of crystallines decrease with increasing temperature proportional to $\kappa(T)\propto 1/T$ known as the Umklapp process\,\cite{Landau1979}.

The theoretical elucidation on the linear rise on temperature ``above" the plateau region has been the main subject of the present paper.
Our calculated results given in Sec.\,V, based on hopping process, show fairly good agreement with observed thermal conductivities above the plateau.
We  particularly emphasize that both the magnitude and the temperature dependence
of $\kappa(T)$ are in accord with the experimental data~\cite{Nolas:1998aa, Cohn:1999a, Sales:2001a, Paschen:2001a, Avila:2006a, Suekuni:2007a, Avila:2008aa, Suekuni:2008a}.
%Thus it has become clear that the T-linear rise is associated with the hopping of SL modes via anharmonic interaction between SL modes and EX acoustic phonons.
At much higher temperatures, the {\it T}-linear rise in $\kappa(T)$ does not continue, but $\kappa(T)$ saturates above $T\simeq$\,100\,K~\cite{Sales:2001a,  Avila:2006a,  Suekuni:2007a, Avila:2008aa}.
In this temperature regime, the treatment based on quantum mechanical process  does not hold for since the life-time of excited modes becomes much smaller than the inverse of their angular frequencies, where
% the guest ions execute actual ``rattling" motion because
the guest ions become free from the constraint of atoms constituting cages.
%The guest ions in cages should behave in a different way from those of atoms in conventional crystalline solids, Under this situation, .
This subject will be discussed in detail elsewhere\,\cite{Nakayama2017}.
%We have given the formula of Eq.\,(\ref{eq:Bridgman2}) to explain observed $\kappa(T)$  of off-center type-I clathrates.
%We have compared theoretical values evaluated  from Eq.\,(\ref{eq:Bridgman2}) with experimental data of $\kappa(T)$  \red{in Figure ???}, which shows the good agreement  with the dependence on the cage sizes.
%The mechanism  is crucial for  thermopower devices since the operating temperatures are mostly higher than room temperature.
%We will publish elsewhere the results of this saturation regime.

In conclusion, the phenomenon of {\it T}-linear rise of $\kappa_L(T)$ above a few 10K in off-center type-I clathrates has been quantitatively explained by analytic theory, on the grounds that off-center clathrates possess definite microscopic structure.
Our successful clarification in quantitative manner is owing to the fact that the systems are more tractable than network-forming glasses with the difficulty to identify relevant constituents at atomistic level caused by their complex microscopic structures.
%\,\red{\cite{Nakayama1999}}.
%This is the main reason that the arguments on thermal conductivities of glasses have continued for decades.
%\begin{figure}[t]
%\includegraphics[width=0.4\linewidth]{state.eps}
%\caption{(Color online) Typical eigenstates of our model in different phases: (a) $\lambda=0.2$, (b, c) $\lambda=1.0$, (d)$\lambda=2.0$, (e, f) $\lambda=3.0$, respectively. Insets are the logarithmic plots of localized states, showing the exponentially decaying tails $\log \left| a_n \right| \propto \exp(-\gamma x)$. The parameters are the same as those in Fig.~\ref{fig1} (b).}\label{fig2} \end{figure}
%\begin{figure}[htb]
%\includegraphics[width=0.3\linewidth]{IPR-c2-v.eps}
%\caption{(Color online) Dependence of IPR values of individual eigenstates on $c_2$ and $\lambda$ for the lattices Eq.~(\ref{eq3}).Shown are the eigenstates labeled by $i$. In the calculations the lattices are kept being self-dual. Parameters: $N=600, \alpha=(\sqrt{5}-1)/2$.}\label{fig3}
%\end{figure}
\\
\\
\textit{Acknowledgments.}~
This work is supported by the National Natural Science Foundation of China Grant No. 11334007 and No. 51506153.
%\red{We are grateful to Zhongwei Zhang and Jie Chen for their  support  to perform first principles calculations to estimate the anharmonic coupling constants.}
J. Z. is supported by the program for Professor of Special Appointment
(Eastern Scholar) at Shanghai Institutions of Higher Learning No. TP2014012.
T. N. acknowledges the support from Grand-in-Aid for Scientific Research from the MEXT in Japan, Grand No.26400381.

\end{document}